\documentclass[prb,twocolumn,superscriptaddress,floatfix,longbibliography]{revtex4-2}
\usepackage{enumerate}
\usepackage{graphicx}
\usepackage{amsmath}
\usepackage{mathtools}
\usepackage{xcolor}
\usepackage{amsfonts}
\usepackage{amssymb}
\usepackage{units}
\usepackage[normalem]{ulem}
\newcommand{\ket}[1]{| #1 \rangle}

\newcommand{\R}[1]{{\color{red} #1}}

\usepackage{dcolumn}
\usepackage{bm}
\usepackage[colorlinks,citecolor=blue,linkcolor=blue,urlcolor=blue]{hyperref}

\begin{document}

\title{Temperature-anisotropy conjugate magnon squeezing in antiferromagnets}
\today
\author{Mahroo Shiranzaei\footnote{Electronic address: mahroo.shiranzaei@physics.uu.se}}
 \affiliation{Division of Materials Theory, Department of Physics and Astronomy, Uppsala University, Box 516, SE-75120 Uppsala, Sweden}
 \author{Jonas Fransson}
 \affiliation{Division of Materials Theory, Department of Physics and Astronomy, Uppsala University, Box 516, SE-75120 Uppsala, Sweden}
\author{Vahid Azimi Mousolou\footnote{Electronic address: v.azimi@sci.ui.ac.ir}}
\affiliation{Division of Materials Theory, Department of Physics and Astronomy, Uppsala University, Box 516, SE-75120 Uppsala, Sweden}
\affiliation{Department of Applied Mathematics and Computer Science, 
Faculty of Mathematics and Statistics, 
University of Isfahan, Isfahan 81746-73441, Iran}

\begin{abstract}
Quantum squeezing is an essential asset in the field of quantum science and technology. In this study, we investigate the impact of temperature and anisotropy on squeezing of quantum fluctuations in two-mode magnon states within uniaxial antiferromagnetic materials. Through our analysis, we discover that the inherent nonlinearity in these bipartite magnon systems gives rise to a conjugate magnon squeezing effect across all energy eigenbasis states, driven by temperature and anisotropy. We show that temperature induces amplitude squeezing, whereas anisotropy leads to phase squeezing. In addition, we observe that the two-mode squeezing characteristic of magnon eigenenergy states is associated with amplitude squeezing. 
This highlights the constructive impact of temperature and the destructive impact of anisotropy on two-mode magnon squeezing. Nonetheless, our analysis shows that the destructive effect of anisotropy is bounded. We demonstrate this by showing that, at a given temperature, the squeezing of the momentum (phase) quadrature (or equivalently, the stretching of the position (amplitude) quadrature) approaches a constant function of anisotropy after a finite value of anisotropy. Moreover, our study demonstrates that higher magnon squeeze factors can be achieved at higher temperatures, smaller levels of anisotropy, and closer to the Brillouin zone center. All these characteristics are specific to low-energy magnons in the uniaxial antiferromagnetic materials that we examine here.  
\end{abstract}

\maketitle

\section{Introduction}

Quantum noise and fluctuations are inherent to a physical system due to the quantum uncertainty principle. Despite the limitation imposed by the Heisenberg uncertainty principle on simultaneous measurements of non-commuting quantum observables with arbitrary precision, the noise of a single quantum observable can be reduced without limitation through quantum squeezing \cite{drummond2004quantum}.

Quantum squeezing has been realized in a variety of systems such as the electromagnetic field \cite{walls1994gj, reid2004quantum, andersen201630}, the vibrational mode in solids and molecules \cite{garrett1997vacuum, garrett1997ultrafast, PhysRevLett.70.3388, RevModPhys.86.1391}, trapped ion \cite{kienzler2015quantum}, magnetic and spin systems \cite{wineland1992spin, PhysRevLett.93.107203, PhysRevB.73.184434, ma2011quantum, PhysRevLett.109.253605, hamley2012spin, PhysRevLett.122.223203, bao2020spin, schnabel2017squeezed, aasi2013enhanced, PhysRevLett.104.250801} . It plays an important role in many applications, for example, it can be used to improve the sensitivity of laser interferometers \cite{schnabel2017squeezed}, increase the accuracy of gravitational wave detection \cite{aasi2013enhanced}, and atomic clocks \cite{PhysRevLett.104.250801, louchet2010entanglement, PhysRevLett.117.143004}. Squeezed states are used to enhance quantum metrology \cite{polino2020photonic} and quantum imaging \cite{moreau2019imaging} tasks. Quantum squeezing is an essential resource to realize continuous variable quantum information processing \cite{RevModPhys.77.513, adesso2014continuous} including protocols for quantum communication \cite{PhysRevLett.109.100502, zhou2017new, zhang2019quantum, srikara2020continuous}, unconditional quantum teleportation \cite{furusawa1998unconditional} and one-way quantum computing \cite{PhysRevLett.101.130501}. Among different classes of Gaussian states, two-mode squeezed states are of particular importance in these applications. Indeed, two-mode squeezed states are commonly produced in the laboratory and are strongly related to quantum entanglement and  Einstein-Podolsky-Rosen nonlocal quantum correlations \cite{PhysRevLett.91.107901}. 

Although the squeezed states were originally discussed in the context of photons, they arise naturally in any bosonic systems, including phonons and magnons. Quantum squeezing in magnonic systems has recently received special attention \cite{yuan2022quantum}. This is related to the fact that magnons allow robust squeezed states in equilibrium, which result from energy minimization, unlike the other bosonic counterparts, where the squeezed states are non-equilibrium in nature and are produced through external forces \cite{kamra2020magnon}.  

In this paper, we study the effect of temperature and anisotropy on two-mode magnon squeezing in antiferromagnetic materials. In a nonlinear treatment, we analyze variations of quantum fluctuations as a function of temperature and anisotropy for antiferromagnetic materials subjected to uniaxial anisotropy. We find a conjugate magnon-squeezing behavior, where temperature-induced squeezing and anisotropy-induced squeezing compete with each other. We demonstrate that temperature plays a constructive role in two-mode magnon squeezing, while anisotropy induces destructive contributions. However, our analysis shows that for a finite value of anisotropy it is possible to achieve high temperature and low energy stabilized magnon squeezing at the proximity of the Brillouin zone center. 

The paper is structured as follows. In Sec. \ref{Two-mode-Magnon-System}, we introduce the antiferromagnetic spin system, perform bosonization up to certain nonlinear terms, and apply mean-field approximation to describe the two-mode magnon system in a quadratic form. In Sec. \ref{Tow-mode-magnon-dispersion-and-states}, magnon dispersion, and two-mode magnon states are obtained. Temperature-anisotropy conjugate magnon squeezing effect is discussed in Sec. \ref{Temperature-anisotropy-conjugate-magnon-squeezing}. The paper ends with a conclusion in Sec. \ref{conclusion}.

\section{Physical System}
\label{Two-mode-Magnon-System}
\subsection{Two-mode magnon system}
\label{Model-System}
The antiferromagnetic Heisenberg Hamiltonian with an easy-axis onsite anisotropy can be expressed by \cite{RevModPhys.90.015005},
\begin{align}\label{Hspin}
H = J \sum_{\langle i, j \rangle} \mathbf{S}_i \cdot \mathbf{S}_j - \sum_{i} \mathcal{K}_z( S^z_i)^2,
\end{align}
where $J>0 $ is the antiferromagnetic Heisenberg exchange coupling and $\mathcal{K}_z> 0$ is the uniaxial anisotropy, which distinguishes the $z$ as the easy axis. Through the bosonization procedure, one can apply the Holstein-Primakoff transformation on AFMs \cite{PhysRev.58.1098},
\begin{eqnarray}
S^z_i &=& S - a^\dag_i a_i ,\hspace{.33cm} 
S^-_i = a^\dag_i \sqrt{2S - a^\dag_i a_i},
\nonumber\\
S^z_j &=& - S + b^\dag_j b_j ,\hspace{0.18cm} 
S^-_j = \sqrt{2S - b^\dag_j b_j} \; b_j,
\label{eq:HP2}
\end{eqnarray}
followed by Taylor expansion in the powers of $1/S$ to derive the effective Hamiltonian of elementary excitations with an arbitrary order of interactions,
\begin{equation}
    H = E^c_{0} + H^{(2)} + H^{(4)}+\cdots.
\end{equation}
The first term is the classical ground-state energy given by,
\begin{eqnarray}
E^{c}_0 =
-N (\mathcal{Z} J/2 + \mathcal{K}_z) S^2,
\end{eqnarray}
with $\mathcal{Z}$ being the coordination number, i.e., the number of nearest neighbors. Below we consider up to fourth-order magnon interactions and focus on,
\begin{equation}
    H = H^{(2)} + H^{(4)},
\end{equation}
where the classical energy contribution $E^c_{0}$ is neglected without loss of generality. In real space, we obtain the quadratic and quartic terms of the Hamiltonian as,
\begin{align}
\begin{split}
    H^{(2)}
    =
    &
    J S \sum_{\langle i,j \rangle} \big[
    \big(
    a^\dag_i a_i + b^\dag_j b_j + a_i b_j + a^\dag_i b^\dag_j
    \big)
    \big]
    \\&
    +
    2 S \sum^{N/2}_{i \in \mathcal{A}}
     \mathcal{K}_z a^\dag_i a_i
    +
    2 S \sum^{N/2}_{j \in \mathcal{B}}
    \mathcal{K}_z \big) b^\dag_j b_j
    \,,
\end{split}
\end{align}
and \cite{10.1088/1367-2630/ac94f0},
\begin{align}\label{eq:Hintdelta}
\begin{split}
H^{(4)}=\frac{-J}{4}\sum_{\langle i, j \rangle} \Big[ &
a^\dag_i a_i b^\dag_j b_j + a_i b^\dag_j b_j b_j 
+ (a\leftrightarrow b)\Big]
\\&
-\frac{\mathcal{K}_z}{2} \sum_{i}\Big[
a^\dag_i a^\dag_i a_i  a_i+(a\leftrightarrow b)\Big]+(h.c.).
\end{split}
\end{align}
By using the Fourier transformations
\begin{align}
\begin{split}
& a_i = \sqrt{\frac{2}{N}} \sum_q e^{i \, \mathbf{q} \cdot \mathbf{r}_i} a_q \; ,
\hspace{0.2cm}
i \in \mathcal{A} \; ,
\\&
b_j = \sqrt{\frac{2}{N}} \sum_q e^{i \, \mathbf{q} \cdot \mathbf{r}_j} b_q \; ,
\hspace{0.2cm}
j \in \mathcal{B} \; ,
\end{split}
\end{align}
the quadratic and quartic Hamiltonians in the crystal momentum space are obtained as,
\begin{equation}\label{nonH}
\begin{split}
H^{(2)} = S \sum_{\bf q} \bigg[ &
\big(\mathcal{Z}J +  2 \mathcal{K}_z \big) \big( a^\dag_{\bf q} a_{\bf q} + b^\dag_{\bf q} b_{\bf q} \big) 
\\&
+ \mathcal{Z}J \gamma_{-{\bf q}} a_{\bf q} b_{-{\bf q}}
+ \mathcal{Z}J \gamma_{\bf q} b^\dag_{-{\bf q}} a^\dag_{\bf q}
\bigg],
\end{split}
\end{equation}
and,
 \begin{equation}\label{eq:Hintdelta}
    \begin{split}
    H^{(4)} = \frac{-1}{N} \sum_{q_1, q_2, q_3, q_4}  & \delta_{q_1+q_2, q_3+q_4} 
    \\&
     \Bigg[ 
    2 \; J \; a^\dag_{q_1} a_{q_3} b^\dag_{q_4} b_{q_2}
    \\&
    + J \; \Big( a_{q_1} b^\dag_{q_2} b_{q_3} b_{q_4} + a^\dag_{q_2} a_{q_3} a_{q_4} b_{q_1} \Big)
    \\&
    +
    \mathcal{K}_{z} \; \Big( a^\dag_{q_1} a_{q_3} a^\dag_{q_2} a_{q_4} + b^\dag_{q_1} b_{q_3} b^\dag_{q_2} b_{q_4} \Big)
    \Bigg]
    \\&
    + (h.c.),
      \end{split}
    \end{equation}
with the lattice structure factor $\gamma_{\bf q} = \mathcal{Z}^{-1}\sum_{i=1}^{z} e^{i {\bf q} \cdot \boldsymbol{\delta}_i}$, in which $\boldsymbol{\delta}_i$ denotes the nearest-neighbor vectors,
Note that although in the forth-order, the antiferromagnetic coupling $J$ only introduces an interaction between the two excitation modes $a$ and $b$ on the opposite sublattices, a uniaxial anisotropy $\mathcal{K}_z$ induces interaction between excitations within each sublattice.

\subsection{Mean-field Hamiltonian}
\label{mean-field-approximation}
The Hamiltonian can be simplified by applying Bogoliubov transformation and mean-field approximation. 
The Bogolioubov transformation \cite{ShenPRB2019,RezendeJAP2019},
\begin{align}\label{BogoTuni}
    \begin{pmatrix}
    a_{\bf q} \\
    b^\dag_{-{\bf q}}
    \end{pmatrix}
    =
    \begin{pmatrix}
    \bar{u}_{\bf q} & -\bar{v}_{\bf q} \\
    -v_{\bf q} & u_{\bf q}
    \end{pmatrix}
    \begin{pmatrix}
    \alpha_{\bf q} \\
    \beta^\dag_{-\bf q}
    \end{pmatrix},
\end{align}
where $|u_{\bf q}|^2 - |v_{\bf q}|^2 = 1$, diagonalizes the quadratic term of the Hamiltonian as,
\begin{align}\label{noninteracting}
{H}^{(2)} = \sum_{\bf q} \epsilon_{\bf q } (\alpha^\dag_{\bf q} \alpha_{\bf q}+\beta^\dag_{-\bf q} \beta_{-\bf q}).
\end{align}
By inserting Eq. \eqref{BogoTuni} into the quadratic Hamiltonian in Eq. \eqref{nonH} with the aim of diagonalization, we obtain the following Bogoliubov coefficients,
\begin{subequations}
\begin{align}\label{bogocoeffuni}
u_{\bf q} &= \cosh\theta_{\bf q} = \sqrt{\frac{\mathcal{Z}J S + 2\, \mathcal{K}_z S + \epsilon_{{\bf q}} }{2\,\epsilon_{{\bf q}}}},
\\
v_{\bf q} &= \sinh\theta_{\bf q}= \sqrt{\frac{\mathcal{Z}J S + 2\, \mathcal{K}_z S - \epsilon_{{\bf q}} }{2\,\epsilon_{{\bf q}}}},
\end{align}
\end{subequations}
and the dispersion relation,
\begin{align}\label{baredisp}
\epsilon_{{\bf q}} = S\sqrt{\big(\mathcal{Z}J + 2 \mathcal{K}_z
\big)^2-\big(\mathcal{Z}J |\gamma_{\bf q}|\big)^2}.
\end{align}

As far as the quadratic Hamiltonian is concerned the bosonic eigenmodes $\alpha$ and $\beta$ represent two polarized magnon modes with opposite chiralities \cite{RezendeJAP2019, PhysRevB.97.020402}, which are separable  up to linear approximation in the Holstein-Primakof transformation \cite{PhysRevB.102.224418, PhysRevB.104.224302}. 
The structure factors for 2D square ($\mathcal{Z}=4$) and hexagonal ($\mathcal{Z}=3$) lattices are, $\gamma_{\bf q}=2\left(\cos(a_cq_x)+\cos(a_cq_y)\right)/\mathcal{Z}$ and $\gamma_{\bf q}=e^{ia_cq_x} \left(1+2 e^{-i 3 a_cq_x/2}\cos(\sqrt{3}a_cq_y/2)\right)/\mathcal{Z}$, respectively, where $a_c$ is the lattice constant. 
In the long-wavelength limit $|{\bm q}| \rightarrow 0$, the structure factors of both hexagonal and square lattices reduce to $\gamma_{\bf q} \simeq 1-a_c^2|{\bm q}|^2/4$, which results in the dispersion relation $\epsilon_\mathbf{q} =S \sqrt{ 4\mathcal{K}_z(\mathcal{Z}J+\mathcal{K}_z)+(a_c |{\bm q}| \mathcal{Z}J)^2 /2}$.

For the quartic Hamiltonian, we consider $q_1 = q_3 = q, \; q_2 = q_4 = q^\prime$ and $q_1 = q_4 = q, \; q_2 = q_3 = q^\prime$, associated with the dominant effect of magnon interactions \cite{tani1968frequency, PhysRev.117.117, PhysRevLett.121.187202}, which imply \cite{10.1088/1367-2630/ac94f0},
    \begin{equation}\label{eq:Hint}
    \begin{split}
    H^{(4)}=\frac{-2}{N} \sum_{q, q^\prime} \Bigg[ &
    J \; a^\dag_q a_q b^\dag_{q^\prime} b_{q^\prime}
    + J \; a^\dag_q a_{q^\prime} b^\dag_q  b_{q^\prime}
    \\&
    + J \; \bigg(a_q b^\dag_{q^\prime} b_{q^\prime} b_q + a^\dag_{q^\prime} a_{q^\prime} a_q b_q \bigg)
    \\&
    +  \frac{\mathcal{K}}{2}
    \bigg( a^\dag_q a_q a^\dag_{q^\prime} a_{q^\prime} + a^\dag_q a_{q^\prime} \; a^\dag_{q^\prime} a_q
    \\& 
    \hspace{1cm} b^\dag_q b_q b^\dag_{q^\prime} b_{q^\prime} + b^\dag_q b_{q^\prime} \; b^\dag_{q^\prime} b_q \bigg)
    \Bigg]
    \\&
    + (\mathit{h.c.}).
      \end{split}
    \end{equation}    
To reduce this quartic Hamiltonian into a mean-field quadratic Hamiltonian, we use Hartree-Fock approximation \cite{balucani1980magnetic, OguchiPR1960} on Eq. \eqref{eq:Hint}. 
Consequently, the finite mean-field contributions come from the following terms,
\begin{eqnarray}
\chi&=& \frac{2}{N S} \sum_{\bf q} \langle a^\dag_{\bf q} a_{\bf q} \rangle = \frac{2}{N S}\sum_{\bf q}  \langle b^\dag_{\bf q} b_{\bf q} \rangle,\nonumber\\
\chi^\prime &=& \frac{2}{N S} \sum_{\bf q} \gamma_{-{\bf q}} \langle a_{\bf q} b_{-{\bf q}} \rangle.
\label{meanfieldpara}
\end{eqnarray}
in momentum space. $\chi$ stands for the number of bosonic excitations on each sublattice $\cal{A}$ and $\cal{B}$, while $\chi^\prime$ denotes the interaction between them. 
By inserting the Bogoliubov transformation given in Eq. \eqref{BogoTuni} into Eq. \eqref{meanfieldpara}, the mean-field parameters in the $(\alpha, \beta)$ magnon modes can be obtained as,
\begin{eqnarray}
\chi&=&\frac{2}{N S} \sum_{\bf q} \big(|u_{\bf q}|^2+|v_{\bf q}|^2\big) \, n_{\bf q} + |v_{\bf q}|^2,
\nonumber\\
\chi^\prime&=& -\frac{2}{N S} \sum_{\bf q} \gamma_{{\bf q}} u_{\bf q} v_{\bf q} \, (2 n_{\bf q} + 1 ),
\label{meanfieldparauni}
\end{eqnarray}
where $n_{\bf q}=(e^{\epsilon_{{\bf q}}/k_B T} - 1)^{-1}$. Here $k_B$ and $T$ are the Boltzmann constant and the temperature respectively. 
Note that in the Hartree-Fock approximation, we only keep the diagonal terms and assume the thermal average $\langle \alpha_{\bf q}^{\dagger} \alpha_{\bf q'}\rangle_{\text{th}}=\langle \beta_{\bf q}^{\dagger} \beta_{\bf q'}\rangle_{\text{th}}=\delta_{{\bf q}{\bf q}'}n_{{\bf q}}$ \cite{LI201828, PhysRevB.104.064435}. 
Therefore, by using the Bogoliubov transformation in Eq. \eqref{BogoTuni} together with the mean-field parameters in Eqs. \eqref{meanfieldpara} and \eqref{meanfieldparauni}, the quartic Hamiltonian in Eq.\ \eqref{eq:Hint} reduces into the following effective quadratic Hamiltonian,
\begin{eqnarray}
{H}^{(4)}=\sum_{\bf q}\tilde{\epsilon}_{\bf q}
(\alpha^{\dagger}_{\bf q}\alpha_{\bf q} + \beta^{\dag}_{-\bf q}\beta_{-\bf q} )
+g_{\bf q}\alpha_{\bf q}\beta_{-\bf q}
+\bar{g}_{\bf q}\beta^\dag_{-\bf q}\alpha^\dag_{\bf q},\nonumber\\
\label{effectivH4uni}
\end{eqnarray}
in $(\alpha, \beta)$ magnon modes with
\begin{eqnarray}
\tilde{\epsilon}_{\bf q} & =& 
[\Lambda (|u_{\bf q}|^2 + |v_{\bf q}|^2) - (\bar{\Lambda^\prime}_{\bf q} \,  \bar{u}_{\bf q} \bar{v}_{\bf q} + \Lambda^\prime_{\bf q} \, u_{\bf q} v_{\bf q})],
\nonumber\\
g_{\bf q}&=&
-2 \Lambda \bar{u}_{\bf q} v_{\bf q} + \Lambda^\prime_{\bf q} (v_{\bf q})^{2} + \bar{\Lambda^\prime}_{\bf q} (\bar{u}_{\bf q})^2,
\label{eq:unimeanH-C}
\end{eqnarray}
where
\begin{eqnarray}
\Lambda_{\bf q} & =&-(\mathcal{Z}J+4\mathcal{K}_z)S\chi-\mathcal{Z}JS\,\text{Re}[\chi^\prime],
\nonumber\\
\Lambda^\prime_{\bf q}&=&-\mathcal{Z}JS(\chi+\bar{\chi^\prime})\gamma_{\bf q}.
\label{eq:unimeanH-C}
\end{eqnarray}

Therefore, the resulting total magnon Hamiltonian of a uniaxial AFM system in the mean-field approximation is
\begin{equation}
\begin{split}
H &=
{H}^{(2)}+{H}^{(4)}
\\&= 
\sum_{\bf q} 
(\epsilon_{\bf q } + \tilde{\epsilon}_{\bf q}) (\alpha^\dag_{\bf q} \alpha_{\bf q} + \beta^\dag_{-\bf q} \beta_{-\bf q})
+ g_{\bf q} \alpha_{\bf q} \beta_{-\bf q} + \bar{g}_{\bf q} \beta^\dag_{-\bf q} \alpha^\dag_{\bf q}.\nonumber
\label{TTH}
\end{split}
\end{equation}
The mean-field contribution $\tilde{\epsilon}_{\bf q}$ renormalize noninteracting magnon modes described by the linear spin-wave Hamiltonian $H^{(2)}$ in Eq.\ \eqref{noninteracting}. Even at zero temperature, these coefficients are finite. Therefore, in general, there is always a finite nonlinear quantum correction to the bare magnon dispersion in AFM systems \cite{10.1088/1367-2630/ac94f0}.

\section{Two-mode magnon dispersion and states}
\label{Tow-mode-magnon-dispersion-and-states}
\subsection{Magnon dispersion}
Due to the presence of interband interaction, $g_{\bf q}$, the Hamiltonian $H$ in Eq.\ \eqref{TTH} is no longer diagonal in the
$(\alpha,\beta)$ modes. To diagonalize $H$, we use the following Bogolioubov transformation,
\begin{eqnarray}
    \begin{pmatrix}
    \alpha_{\bf q} \\
    \beta^\dag_{-{\bf q}}
    \end{pmatrix}
    =
    \begin{pmatrix}
    \tilde{u}_{\bf q} & -e^{-i\phi_{\bf q}}\tilde{v}_{\bf q} \\ 
   -e^{i\phi_{\bf q}}\tilde{v}_{\bf q} & \tilde{u}_{\bf q}
    \end{pmatrix}
    \begin{pmatrix}
    \eta_{\bf q} \\
    \zeta^\dag_{-\bf q}
    \end{pmatrix},
\label{BT2}
\end{eqnarray}
with the parameters \cite{PhysRevB.104.224302},
\begin{eqnarray}
\tilde{u}_{\bf q} &=& \cosh\tilde{\theta}_{\bf q} = \sqrt{\frac{\epsilon_{\bf q} + \tilde{\epsilon}_{\bf q} + \mathcal{E}_{\bf q}}{2\,\mathcal{E}_{\bf q}}},
\nonumber\\
\tilde{v}_{\bf q} &=& \sinh\tilde{\theta}_{\bf q} = \sqrt{\frac{\epsilon_{\bf q} + \tilde{\epsilon}_{\bf q} - \mathcal{E}_{\bf q}}{2\,\mathcal{E}_{\bf q}}},
\\
\phi_{\bf q} &=&\arg[g_{\bf q}],\ \ \ \ \ \ \mathcal{E}_{\bf q}=
\sqrt{\left(\epsilon_{\bf q} + \tilde{\epsilon}_{\bf q}\right)^2-|g_{{\bf q}}|^2}.\nonumber
\label{bogocoeffuni2}
\end{eqnarray}

\begin{figure}[h]
    \centering
    \includegraphics[width=0.5\columnwidth]{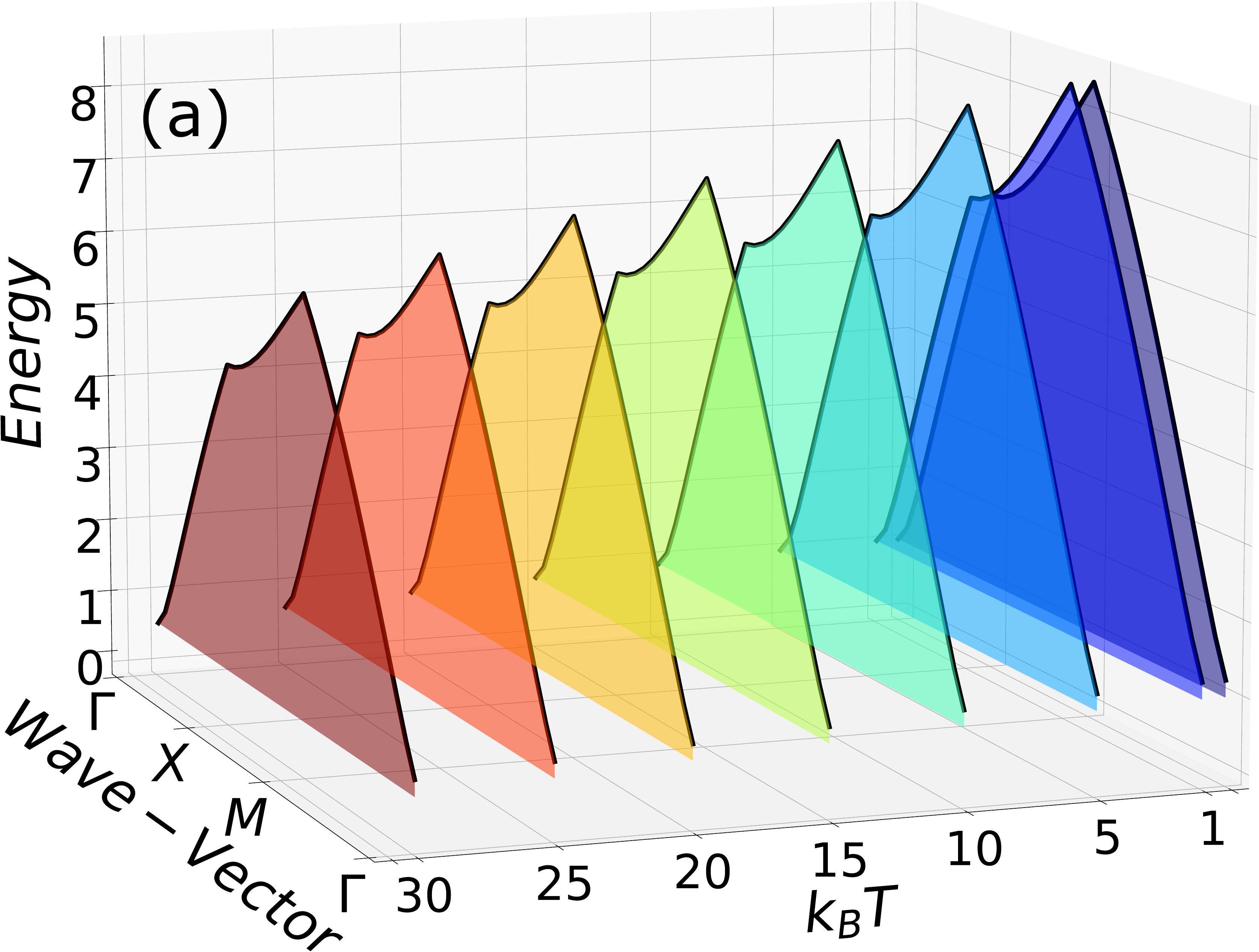}
    {\includegraphics[width=0.45\columnwidth]{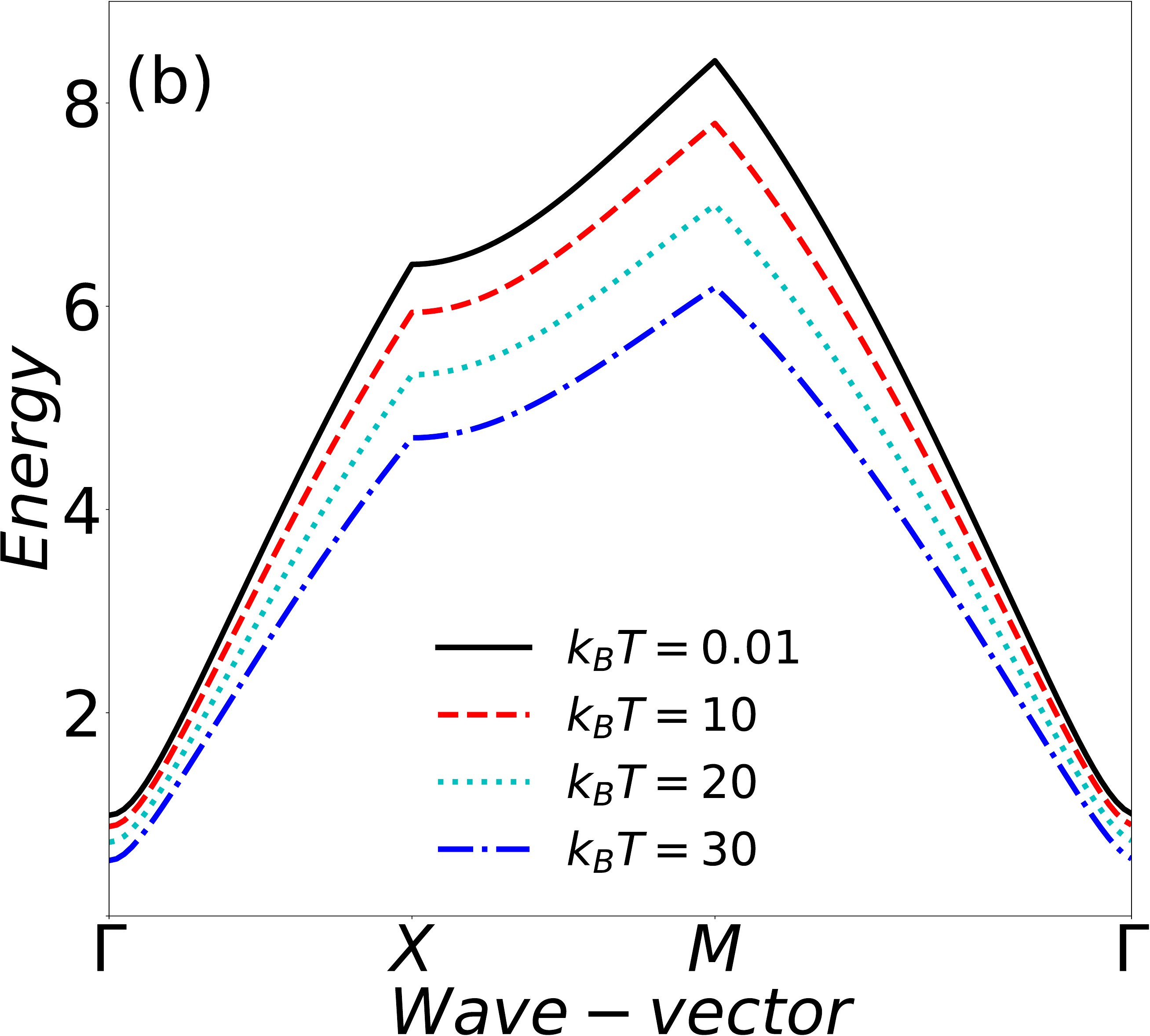}}
    \caption{Magnon dispersion of a bipartite antiferromagnet as a function of temperature and wave vector. We consider a square lattice with an easy-axis anisotropy $\mathcal{K}_z = 0.01$ meV. Different cross sections in panel (a) and hence different lines in panel (b) correspond to different values of temperatures. The energy decreases with temperature and the system is more stable against thermal fluctuations at the center of the Brillouin zone.}
    \label{fig:dispersion1}
\end{figure}

\begin{figure}[h]
    \centering
    {\includegraphics[width=0.5\columnwidth]{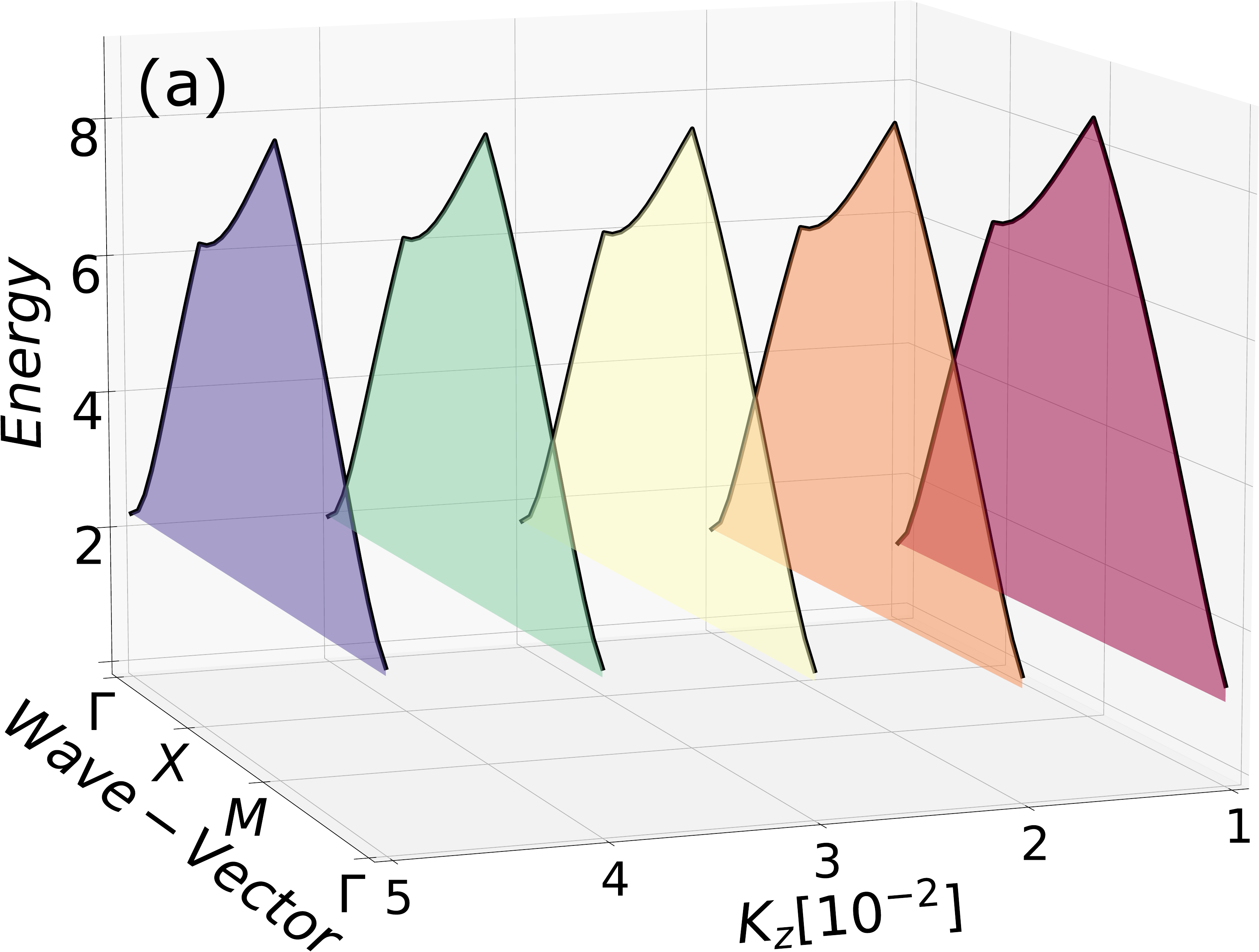}}
    {\includegraphics[width=0.45\columnwidth]{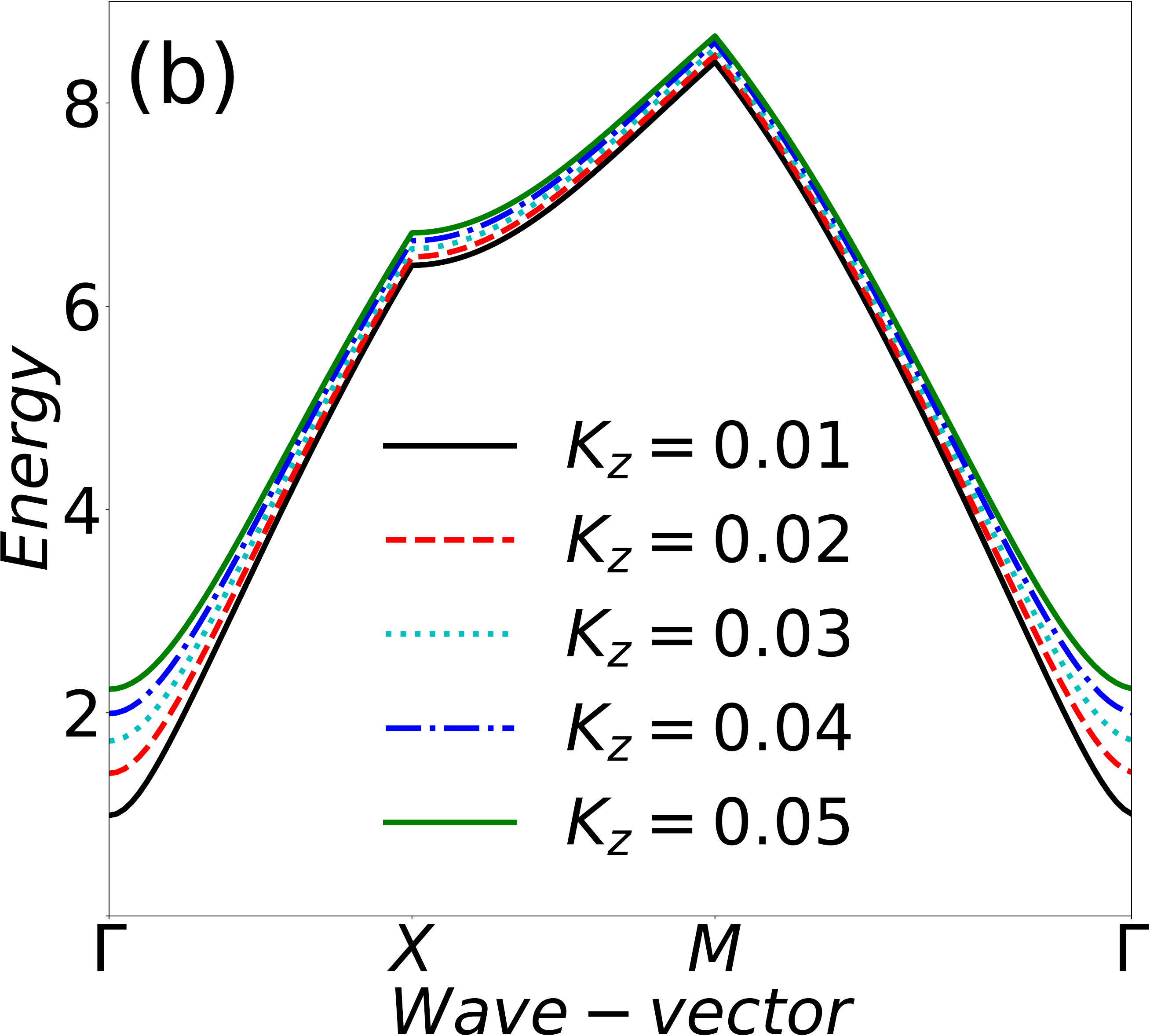}}
    \caption{Magnon dispersion of a bipartite antiferromagnet as a function of easy-axis anisotropy and wave vector. We consider a square lattice at temperature $k_BT = 1$ meV. Different cross sections in panel (a) and hence different lines in panel (b) correspond to different values of anisotropy. The energy increases with anisotropy and the change is relatively more pronounced around the center of the Brillouin zone.}
    \label{fig:dispersion2}
\end{figure}

In the hybridized bosonic modes $(\eta, \zeta)$, the Hamiltonian takes the following diagonal form 
\begin{eqnarray}
H=\sum_{\bf q}\mathcal{E}_{\bf q}\left(\eta^{\dagger}_{\bf q}\eta_{\bf q}+\zeta^{\dagger}_{-\bf q}\zeta_{-\bf q}\right)
\label{DMH}
\end{eqnarray}
with dispersion relation,
\begin{eqnarray}
\mathcal{E}_{\bf q}=
\sqrt{\left(\epsilon_{\bf q} + \tilde{\epsilon}_{\bf q}\right)^2-|g_{{\bf q}}|^2}.
\label{eq.eigen-ene-uni}
\end{eqnarray}
This dispersion is valid for any uniaxial AFMs with arbitrary dimensions and lattice structures at finite temperatures. The evaluation of the dispersion $\mathcal{E}_{\bf q}$, which involves mean-field coefficients through the calculation of $\Lambda$ and $\Lambda^\prime_q$ in Eq. \eqref{eq:unimeanH-C}, is done self-consistently for various temperatures. 
Figs. \ref{fig:dispersion1} and \ref{fig:dispersion2} illustrate the dispersion along a high-symmetric path in the Brillouin zone respectively in terms of different values of temperature and uniaxial anisotropy for an antiferromagnet on a square lattice. Our analysis shows that while the magnon energy dispersion is lower at higher temperatures, a uniaxial anisotropy increases the magnon energy dispersion. In a word, the lower anisotropy is favorable to low-energy magnons at high temperatures.    

From a practical point of view, the low-energy and high-temperature magnon regime is promising for sustainable quantum development. To confirm this, below we investigate how the magnon squeezing behaves in this regime.

\subsection{Two-mode Magnon states}
\label{Two-mode-Magnon-states}
From Eq. \eqref{DMH}, we obtain the magnon energy eigenbasis states to be the 
the following magnon occupation number basis
\begin{eqnarray}
\ket{\psi^{{\bf q}}_{nm}}=\ket{n; \eta_{\bf q}}\ket{m; \zeta_{-\bf q}}=\left(\eta^{\dagger}_{\bf q}\right)^{n}\left(\zeta^{\dagger}_{\bf q}\right)^{m}\ket{0; \eta_{\bf q}}\ket{0; \zeta_{-\bf q}}\nonumber\\
\label{LPMANS}
\end{eqnarray}
in the hybridized $(\eta, \zeta)$ magnon modes, where $\ket{\psi^{{\bf q}}_{00}}=\ket{0; \alpha_{\bf q}}\ket{0; \beta_{-\bf q}}$ is the magnon vacuum ground state. From now on we focus on single ${\bf q}$ vector as different ${\bf q}$s are decoupled in the Hamiltonian of Eq. \eqref{DMH}.

\begin{figure}[h]
    \centering
    \includegraphics[width=0.48\columnwidth]{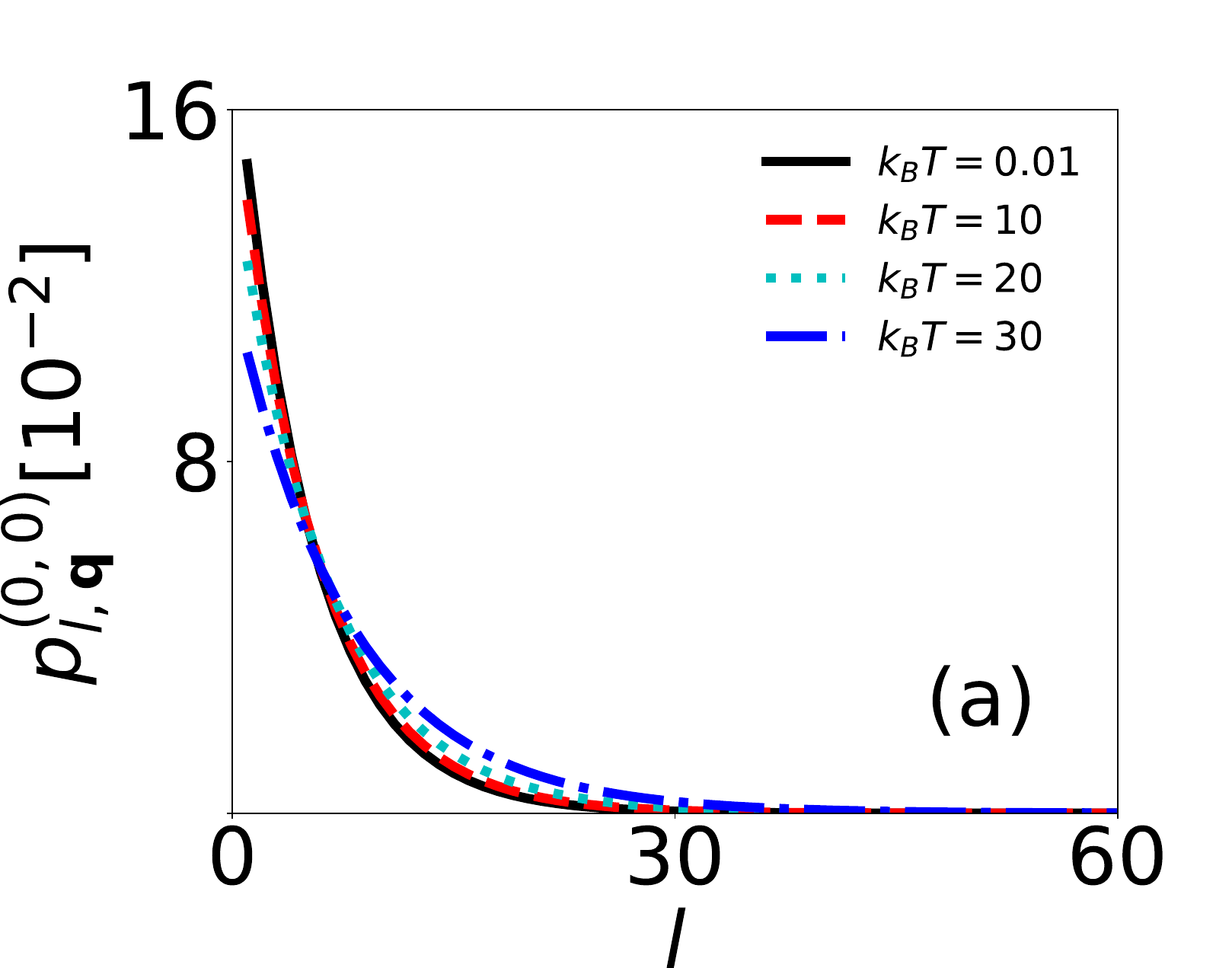}
    {\includegraphics[width=0.48\columnwidth]{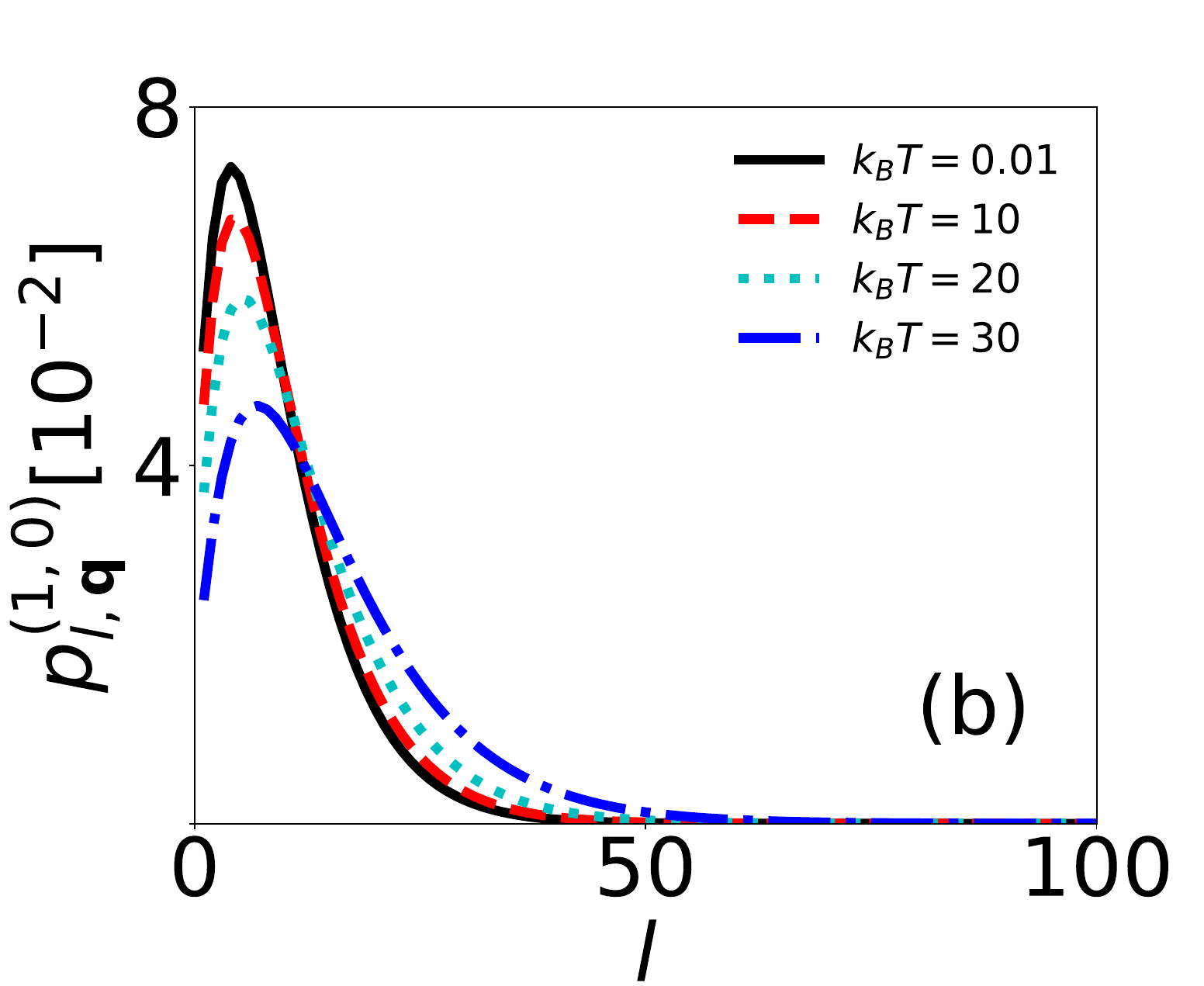}}
    \includegraphics[width=0.48\columnwidth]{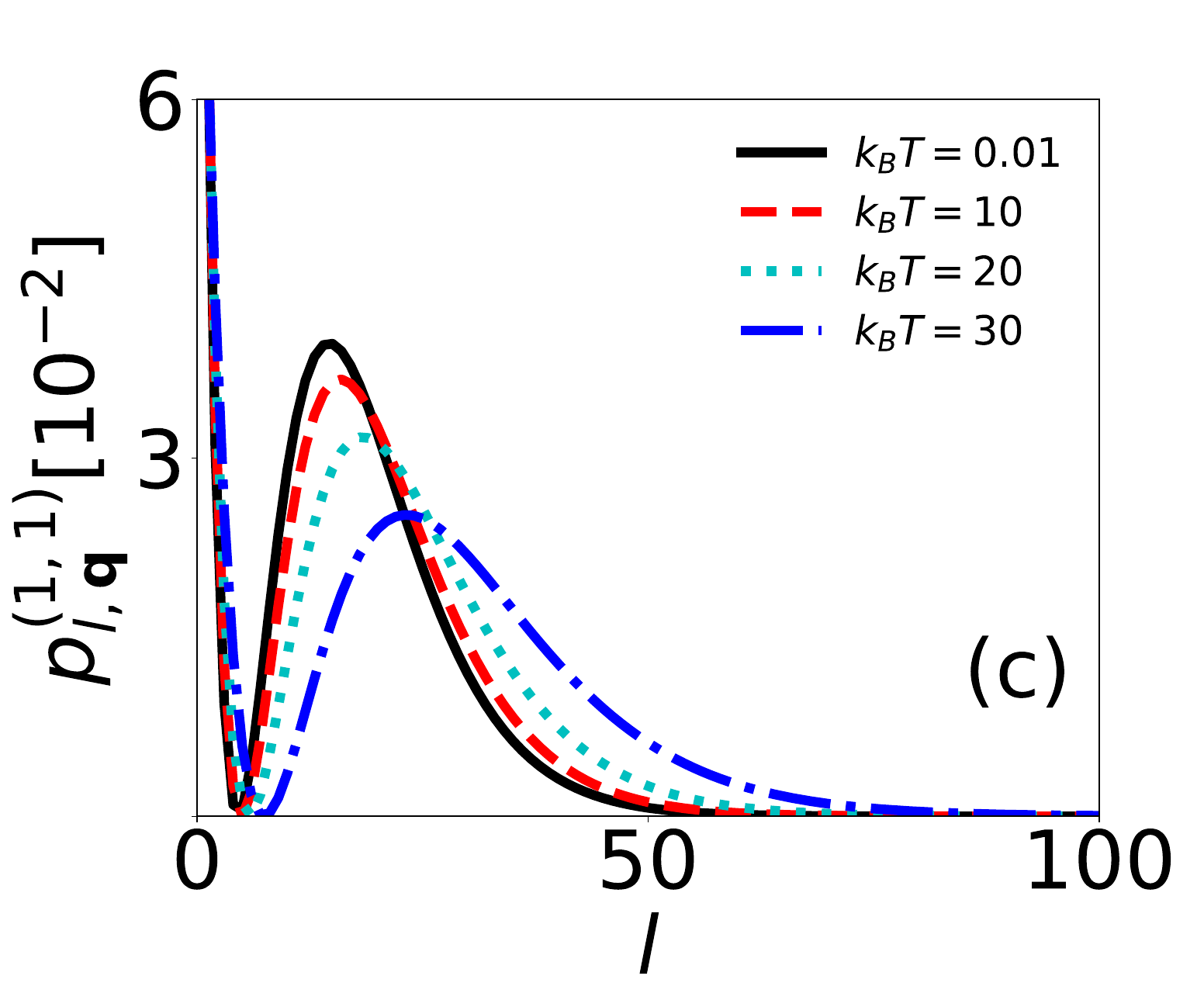}
    {\includegraphics[width=0.48\columnwidth]{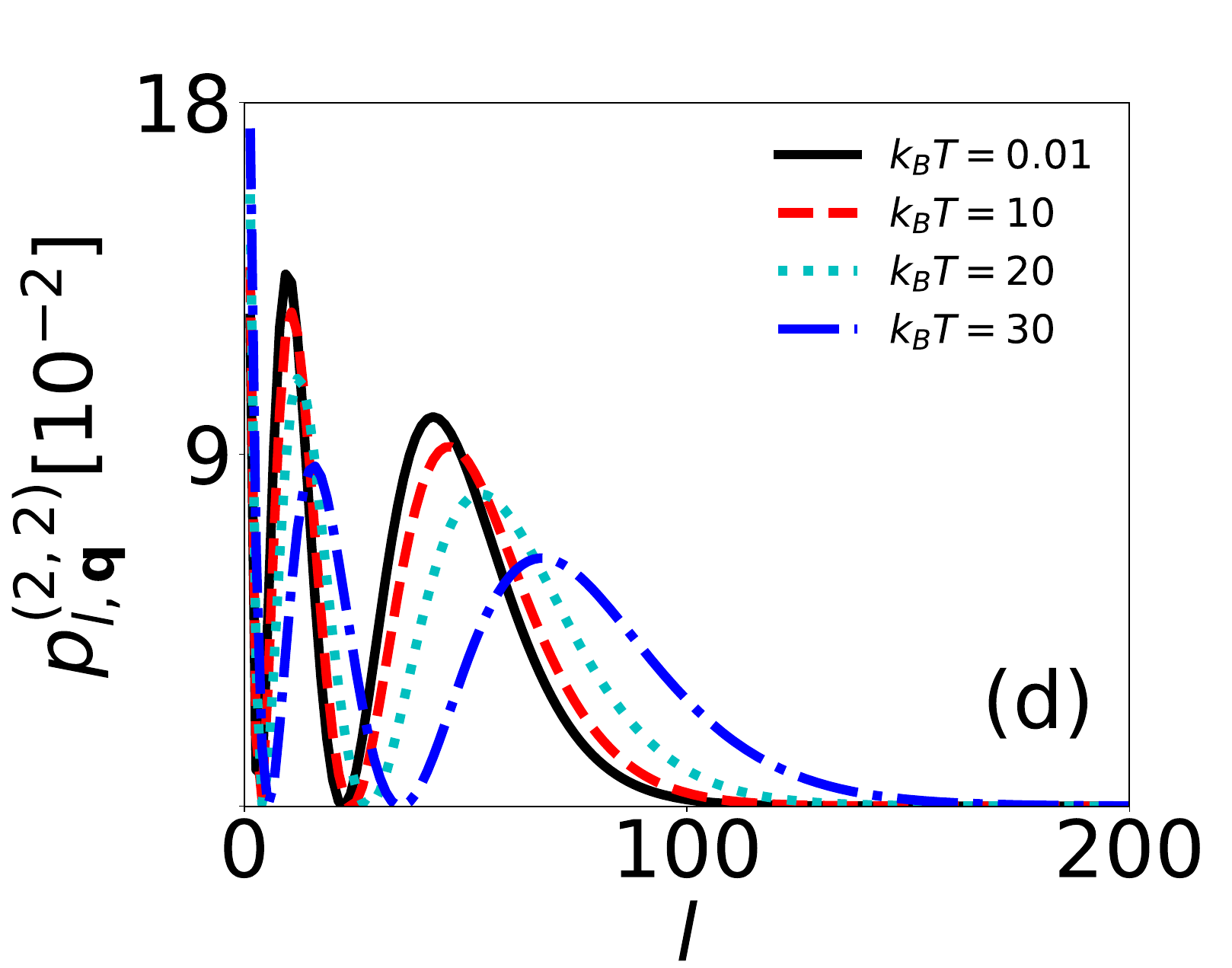}}
    \caption{The probability amplitudes of the coherent expansions of 
    energy eigenbasis states in the Kittel modes $(a, b)$ 
    as a function of the occupation number $l$ at the zone center $\mathbf{q}=0$. We consider a square lattice with an easy-axis anisotropy $\mathcal{K}_z = 0.01$ meV. The probability amplitudes are shown for vacuum state in panel (a) and for a few excited states in panels (b-d). In general, the probability amplitudes tend to diminish for large occupation numbers.}
    \label{fig:probability}
\end{figure}

Here, we are interested in the Kittel magnon modes $(a, b)$, which naturally describe identifiable elementary excitation modes being associated with each sublattice of the AFM through the Holstein-Primakoff transformation in Eq. \eqref{eq:HP2}. The Kittel magnon modes are related to the hybridized modes $(\eta, \zeta)$ through the transformation,
\begin{eqnarray}
    \begin{pmatrix}
    a_{\bf q} \\
    b^\dag_{-{\bf q}}
    \end{pmatrix}
    =
    \begin{pmatrix}
    w_{\bf q} & \nu_{\bf q} \\ 
   \bar{\nu}_{\bf q} & \bar{w}_{\bf q}
    \end{pmatrix}
    \begin{pmatrix}
    \eta_{\bf q} \\
    \zeta^\dag_{-\bf q}
    \end{pmatrix},
\label{BT3}
\end{eqnarray}
where
\begin{eqnarray}
    \begin{pmatrix}
    w_{\bf q} & \nu_{\bf q} \\ 
   \bar{\nu}_{\bf q} & \bar{w}_{\bf q}
    \end{pmatrix}
   =
   \begin{pmatrix}
    \bar{u}_{\bf q} & -\bar{v}_{\bf q} \\
    -v_{\bf q} & u_{\bf q}
    \end{pmatrix}
     \begin{pmatrix}
    \tilde{u}_{\bf q} & -e^{-i\phi_{\bf q}}\tilde{v}_{\bf q} \\ 
   -e^{i\phi_{\bf q}}\tilde{v}_{\bf q} & \tilde{u}_{\bf q}
    \end{pmatrix}\ \ \ 
\label{BT22}.
\end{eqnarray}
This follows from the Bogolioubov transformations given in Eqs. \eqref{BogoTuni} \R{and \eqref{BT2}}. By using this relation, the magnon energy eigenbasis states of the total Hamiltonian in the Kittel modes $(a, b)$ take the following coherent form 
\begin{eqnarray}
\ket{\psi^{{\bf q}}_{nm}}&=&\ket{n; \eta_{\bf q}}\ket{m; \zeta_{-\bf q}}\nonumber\\
&\equiv&\left\{ \begin{array}{ll} 
\sum_{l=0}^{\infty} p^{(n, m)}_{l; \bf q}\ket{l+\delta; a_{\bf q}} 
\ket{l; b_{-\bf q}}\ \ \ \ \ \ \ n\ge m \\ 
\sum_{l=0}^{\infty} p^{(n, m)}_{l; \bf q}\ket{l; a_{\bf q}} 
\ket{l+\delta; b_{-\bf q}}\ \ \ \ \ \ \  n\le m\\ 
\end{array} \right. 
\nonumber\\
\label{EES}
\end{eqnarray}
where $\delta=|n-m|$. By induction, we obtain the probability amplitudes
\begin{eqnarray}
p^{(n, m)}_{l; \bf q}=\frac{1}{\sqrt{n!m!}}
\left(\frac{1}{\bar{w}_{\bf q}}\right)^{\delta}\left(\frac{1}{\bar{w}_{\bf q}\nu_{\bf q}}\right)^{\mu}f^{(\mu, \delta)}_{l; \bf q}p^{(0, 0)}_{l; \bf q},\ \ \ 
\end{eqnarray}
where $\mu=\min\{n, m\}$ and
\begin{eqnarray}
p^{(0, 0)}_{l; \bf q}=\frac{e^{il\varphi_{\bf q}}}{\cosh 
r_{\bf q}}\tanh^{l}r_{\bf q}
\label{EXCC}
\end{eqnarray}
with
\begin{eqnarray}
e^{i\varphi_{\bf q}}\tanh r_{\bf q}=\frac{\nu_{\bf q}}{w_{\bf q}}
\end{eqnarray}
being the expansion coefficient of the ground state $\ket{\psi^{{\bf q}}_{00}}$ of the total Hamiltonian in the Kittel modes.
The $f^{(\mu, \delta)}_{l; \bf q}$ satisfies the following recursive relations,
\begin{eqnarray}\label{flq}
f^{(\mu, \delta>0)}_{l; \bf q}&=&\sqrt{l+\delta} \cosh^{2} r_{\bf q}f^{(\mu, \delta-1)}_{l; \bf q}
\nonumber \\  & & - \sqrt{l+1} \sinh^{2} r_{\bf q}f^{(\mu, \delta-1)}_{l+1; \bf q} , \nonumber\\
f^{(\mu>0, 0)}_{l; \bf q}&=&l\cosh^{4} r_{\bf q}f^{(\mu-1, 0)}_{l-1; \bf q}-\frac{(2l+1)}{2}\sinh^{2} 2r_{\bf q}f^{(\mu-1,0)}_{l; \bf q}\nonumber\\
&&+(l+1)\sinh^{2} r_{\bf q}f^{(\mu-1,0)}_{l+1; \bf q},
\end{eqnarray}
with initial value condition $f^{(0, 0)}_{l; \bf q}=1$ for each $l$. Fig. \ref{fig:probability} shows the behavior of probability amplitude with respect to the occupation number $l$ for a few energy eigenbasis states. Eqs. \eqref{EES}-\eqref{flq} follow from the diagonal form of the total Hamiltonian in Eq.\ \eqref{DMH} and the Bogolioubov transformation in Eq.\ \eqref{BT3}.

We note that the ground state 
\begin{eqnarray}
\ket{\psi^{{\bf q}}_{00}}= \frac{1}{\cosh r_{\bf q}}
\sum_{l=0}^{\infty}e^{il\phi_{\bf q}}\tanh^{l}r_{\bf q}.\ket{l; a_{\bf q}}\ket{l; b_{-\bf q}}
\end{eqnarray}
is indeed the two-mode squeezed state, which plays an important role in quantum information \cite{PhysRevLett.91.107901}.

\section{Temperature-anisotropy conjugate magnon squeezing}
\label{Temperature-anisotropy-conjugate-magnon-squeezing}
\begin{figure}[h]
    \centering
    {\includegraphics[width=0.8\columnwidth]{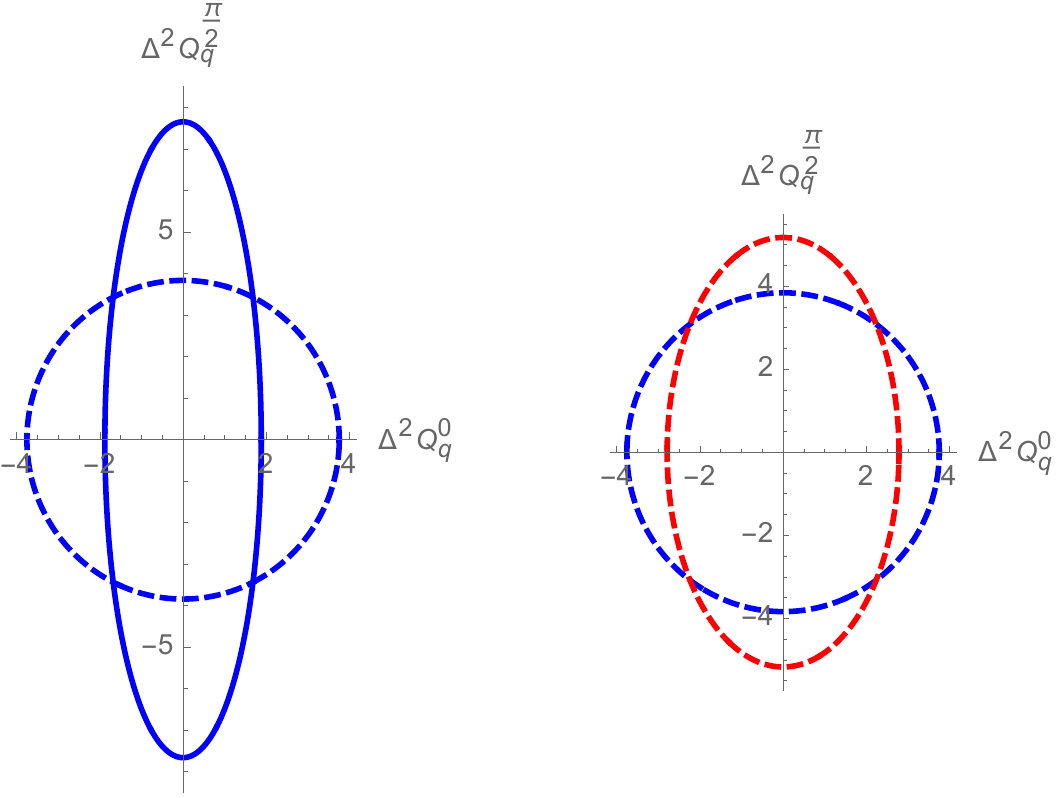}}
    \caption{Schematic illustration of temperature-anisotropy induced conjugate magnon squeezing. Red (blue) color in the right panel indicates higher (lower) temperature for a fixed value of anisotropy. This shows that increasing temperature induces squeezing (antisqueezing) in the quadrature, $Q_{\bf q}^{0}$ (the conjugate quadrature, $Q_{\bf q}^{\pi/2}$).
    Dashed (solid) ellipse in the left panel corresponds to higher (lower) uniaxial anisotropy at a fixed temperature.
    This exhibit squeezing (antisqueezing) in the conjugate quadrature, $Q_{\bf q}^{\pi/2}$, (the quadrature, $Q_{\bf q}^{0}$,) as a result of increasing the anisotropy at a fixed temperature.}
    \label{fig:T-Q-SQ}
\end{figure}

In this section we aim to explore the two-mode magnon squeezing induced by the temperature, $T$, and the uniaxial anisotropy, $\mathcal{K}_z$, in the AFM system described above. Explicitly, for the two-mode magnon states obtained in Sec. \ref{Two-mode-Magnon-states}, we analyze how the parameters $T$ and $\mathcal{K}_z$ contribute to the reduction of quantum noise in an observable. An interesting nontrivial finding is that temperature and uniaxial anisotropy result conjugate two-mode magnon squeezing effect. 
That means $T$ and $\mathcal{K}_z$ induce squeezing (anti-squeezing) of quantum noise in two distinct conjugate observables. Moreover, as schematically illustrated in Fig. \ref{fig:T-Q-SQ}, 
if the temperature induces squeezing for one observable then the anisotropy shows anti-squeezing for the same observable, and the reverse occurs for the associate conjugate observable.  

To clarify $T/\mathcal{K}_z$ features of two-mode magnon squeezing, we consider the quadrature operator

\begin{eqnarray}
Q_{\bf q}^{\nu}= \frac{1}{\sqrt{2}}\left[e^{-i\nu}\Sigma_{\bf q}+e^{i\nu}\Sigma_{\bf q}^{\dagger}\right]=\cos(\nu)X_{\bf q}+\sin(\nu)P_{\bf q}\nonumber\\
\label{QG}
\end{eqnarray}
at polar angle $\nu\in[0,\pi]$. 
$Q_{\bf q}^{n\pi}\equiv X_{\bf q}$ and $Q_{\bf q}^{(2n+1)\pi/2}\equiv P_{\bf q}$ are the normalized dimensionless position and momentum quadratures respectively,
which are given by $X_{\bf q}=\frac{\Sigma_{\bf q}+\Sigma_{\bf q}^{\dagger}}{\sqrt{2}}$ and $P_{\bf q}=\frac{\Sigma_{\bf q}-\Sigma_{\bf q}^{\dagger}}{i\sqrt{2}}$
in terms of the normalized total annihilation, $\Sigma_{\bf q}=\frac{a_{\bf q}+b_{-\bf q}}{\sqrt{2}}$, and creation, $\Sigma_{\bf q}^{\dagger}=\frac{a_{\bf q}^{\dagger}+b_{-\bf q}^{\dagger}}{\sqrt{2}}$, operators.
These quadratures are conjugate observables, which satisfy the canonical commutation relation $\left[X_{\bf q}, P_{\bf q}\right]=i$ and thus the following uncertainty relation 
\begin{eqnarray}
\Delta^{2} X_{\bf q}\Delta^{2} P_{\bf q}\geq \frac{1}{4}
\label{eq:HUC}
\end{eqnarray}
for quantum fluctuations evaluated by the variance $\Delta^{2}\mathcal{O}=\langle\mathcal{O}^{2}\rangle-\langle\mathcal{O}\rangle^{2}$, $\mathcal{O}=X_{\bf q}, P_{\bf q}$ for each quantum state. In quantum optics, $X_{\bf q}$ and $P_{\bf q}$ are conventionally known as amplitude and phase quadratures, respectively. Note that the uncertainty relation in Eq. \eqref{eq:HUC}, indicates that quantum fluctuations in two conjugate observables $X_{\bf q}$ and $P_{\bf q}$ can not be reduced simultaneously 
below $1/2$. In other words, amplitude squeezing and phase squeezing are not simultaneously possible \cite{drummond2004quantum}. 
In fact, amplitude squeezing causes antisqueezing in the phase quadrature and vice versa. 

For each magnon energy eigenbasis state given in Eq. \eqref{EES}, the variance of the quadrature operator in Eq. \eqref{QG} reads
\begin{eqnarray}
\Delta^{2}Q_{\bf q}^{\nu}=\cos^{2}(\nu)\Delta^{2}X_{\bf q}+\sin^{2}(\nu)\Delta^{2} P_{\bf q},
\end{eqnarray}
where we obtain 
\begin{eqnarray}
4\Delta^{2}X_{\bf q}(n, m)&=&\lvert p^{(n, m)}_{0; \bf q}\sqrt{\delta}\rvert^{2}+\nonumber\\
&&\sum_{l=0}^{\infty}\lvert p^{(n, m)}_{l+1; \bf q}\sqrt{l+\delta+1}+p^{(n, m)}_{l; \bf q}\sqrt{l+1}\rvert^{2}+\nonumber\\
&&\sum_{l=0}^{\infty}\lvert p^{(n, m)}_{l+1; \bf q}\sqrt{l+1}+p^{(n, m)}_{l; \bf q}\sqrt{l+\delta+1}\rvert^{2},
\nonumber\\
4\Delta^{2}P_{\bf q}(n, m)&=&\lvert p^{(n, m)}_{0; \bf q}\sqrt{\delta}\rvert^{2}+\nonumber\\
&&\sum_{l=0}^{\infty}\lvert p^{(n, m)}_{l+1; \bf q}\sqrt{l+\delta+1}-p^{(n, m)}_{l; \bf q}\sqrt{l+1}\rvert ^{2}+\nonumber\\
&&\sum_{l=0}^{\infty}\lvert p^{(n, m)}_{l+1; \bf q}\sqrt{l+1}-p^{(n, m)}_{l; \bf q}\sqrt{l+\delta+1}\rvert^{2}.
\nonumber\\
\end{eqnarray}

To pursue, we first focus on the zone center, where the low-energy
magnons are available (see Fig. \ref{fig:dispersion1} and Fig. \ref{fig:dispersion2}), and the system allows stronger quantum correlation between the two magnons in the kittel modes $(a, b)$ \cite{PhysRevB.102.224418, PhysRevB.104.224302}. Moreover, we continue the main discussion with the vacuum ground state $\ket{\psi^{{\bf q}}_{00}}$ and then extend it to excited magnon states in the Appendix (Sec. \ref{Appendix}). Thus, we drop $n$ and $m$ in what follows for simplicity.

\begin{figure}[h]
\begin{center}
    {\includegraphics[width=0.48\columnwidth]{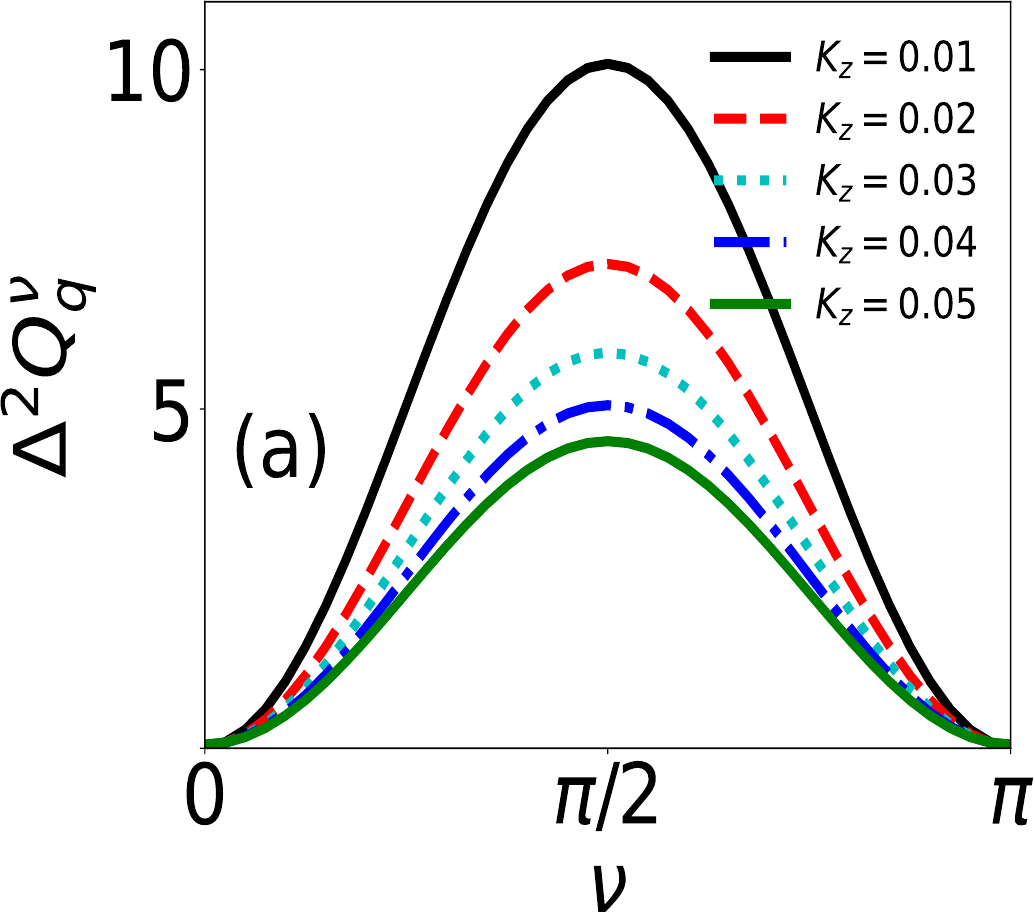}} 
    {\includegraphics[width=0.49\columnwidth]{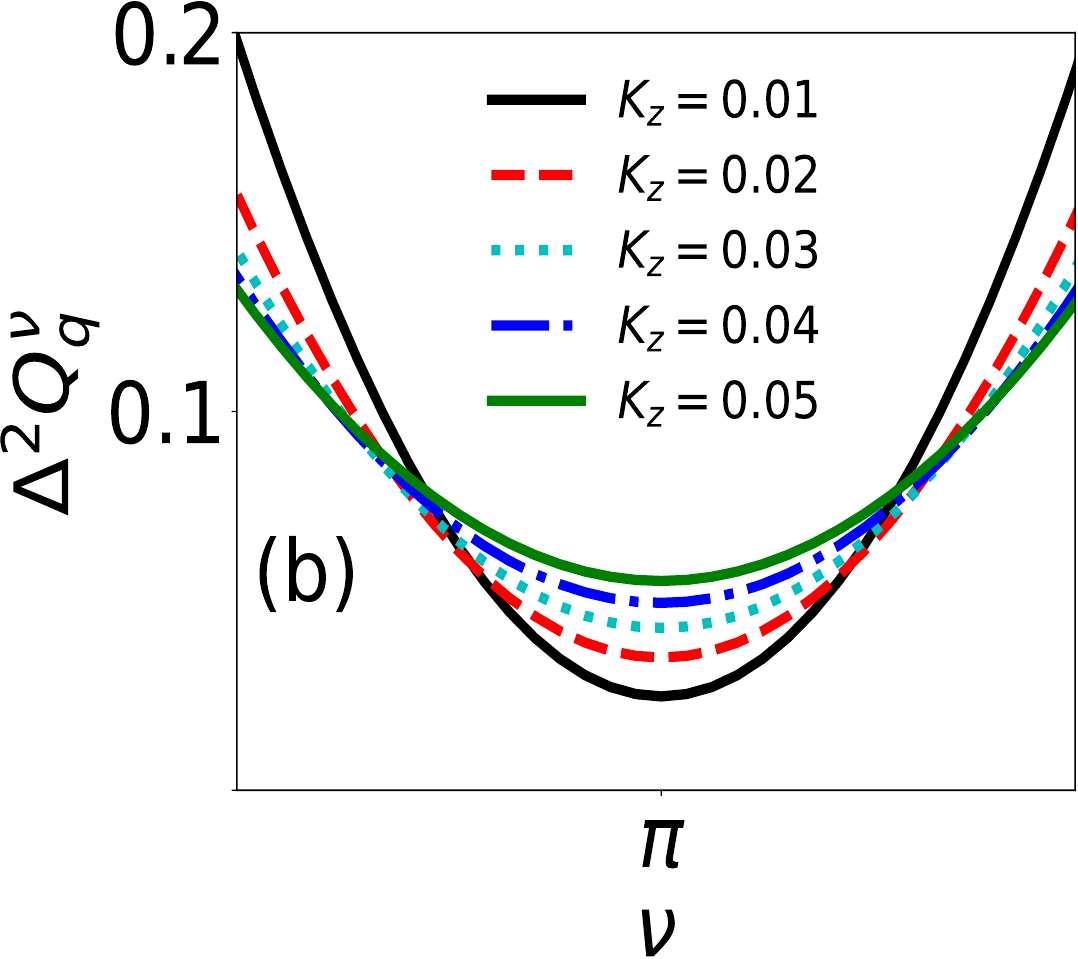}}
    {\includegraphics[width=0.49\columnwidth]{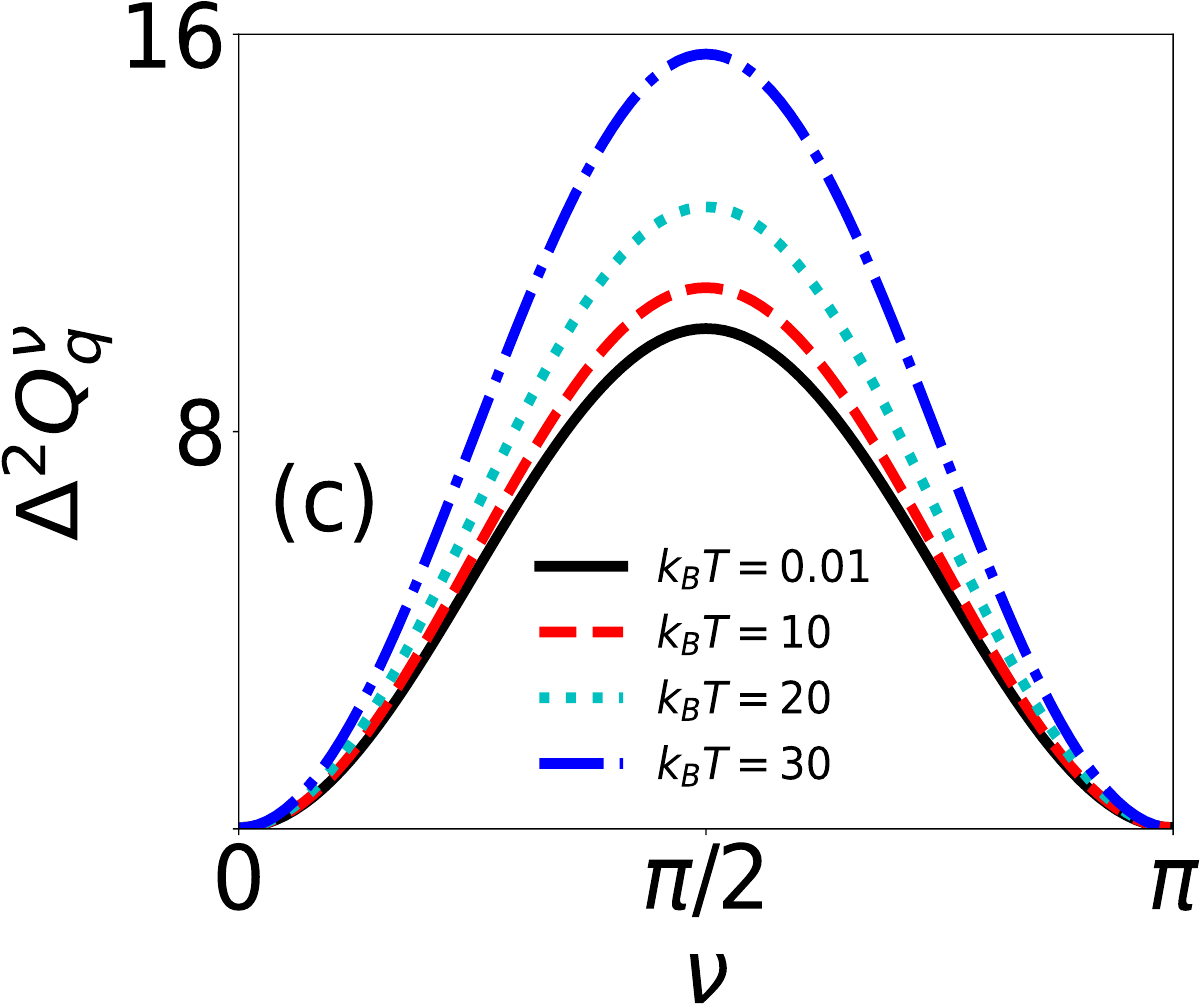}}
    {\includegraphics[width=0.46\columnwidth]{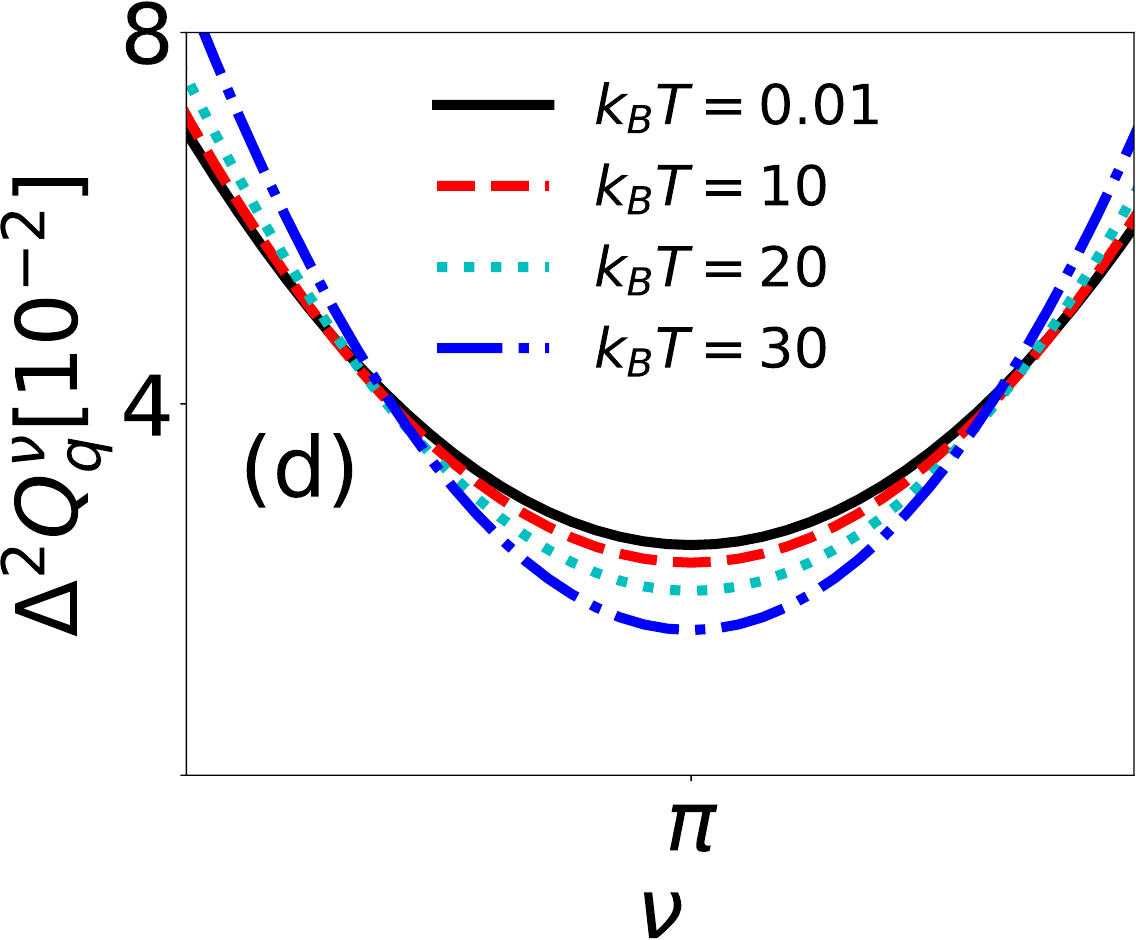}}
\end{center}
\caption{Upper panels: $\Delta^{2}Q_{\bf q}^{\nu}$ in terms of $\nu$ for different easy-axis anisotropies and $k_BT = 1$. Lower panels: $\Delta^{2}Q_{\bf q}^{\nu}$ in terms of $\nu$ for different temperatures and $\mathcal{K}_z =0.01$ meV. 
The focus is $\Gamma$-point which corresponds to ${\bf q}=0$. Temperature and anisotropy induce quantum squeezings at conjugate polar angles, namely, integer and half-integer multiples of $\pi$, respectively. In each of the conjugate angles, if temperature leads to squeezing, anisotropy give rise to antisqueezing of quantum fluctuations and vice versa.}
\label{fig:SPAQ}
\end{figure}
For the vacuum ground state $\ket{\psi_{00}^{\mathbf{q}}}$, Fig.\ \ref{fig:SPAQ} depicts the quantum fluctuation of the quadrature operator, $\Delta^{2}Q_{\mathbf{q}}^{\nu}$, at the zone center, $\mathbf{q}=0$, as a function of the polar angle $\nu$, considering different values of temperature and anisotropy. The impact of temperature and anisotropy on quantum fluctuations is particularly notable when $\nu$ takes on the values of "integer multiples of $\pi$" and "half-integer multiples of $\pi$". These specific values correspond to fluctuations in position (amplitude), denoted as $X_{\mathbf{q}}$, and momentum (phase), denoted as $P_{\mathbf{q}}$, which are conjugate observables. The narrow fluctuation $\Delta^{2}Q_{\bf q}^{\pi}=\Delta^{2}X_{\mathbf{q}}<1/2$, regardless of temperature and anisotropy values, confirms the vacuum ground state $\ket{\psi_{00}^{\mathbf{q}}}$ as a squeezed magnon state. The squeezed nature of the vacuum ground state is, in fact, linked to the squeezing of quantum fluctuations in the position quadrature observable. 
Note that the inequality relation for the variance of the position quadrature saturates, i.e., $\Delta^{2}X_{\mathbf{q}} = 1/2$, if $r_{\mathbf{q}} = 0$. In this case, the vacuum state $\ket{\psi_{00}^{\mathbf{q}}}$ of the $(\eta, \zeta)$ modes coincides with the vacuum state $\ket{0; a_{\mathbf{q}}}\ket{0; b_{-\mathbf{q}}}$ of the Kittel modes $(a, b)$. For the Kittel vacuum state $\ket{0; a_{\mathbf{q}}}\ket{0; b_{-\mathbf{q}}}$, we also find that $\Delta^{2}P_{\mathbf{q}} = 1/2$. 

\begin{figure}[h]
    \centering
    {\includegraphics[width=0.8\columnwidth]{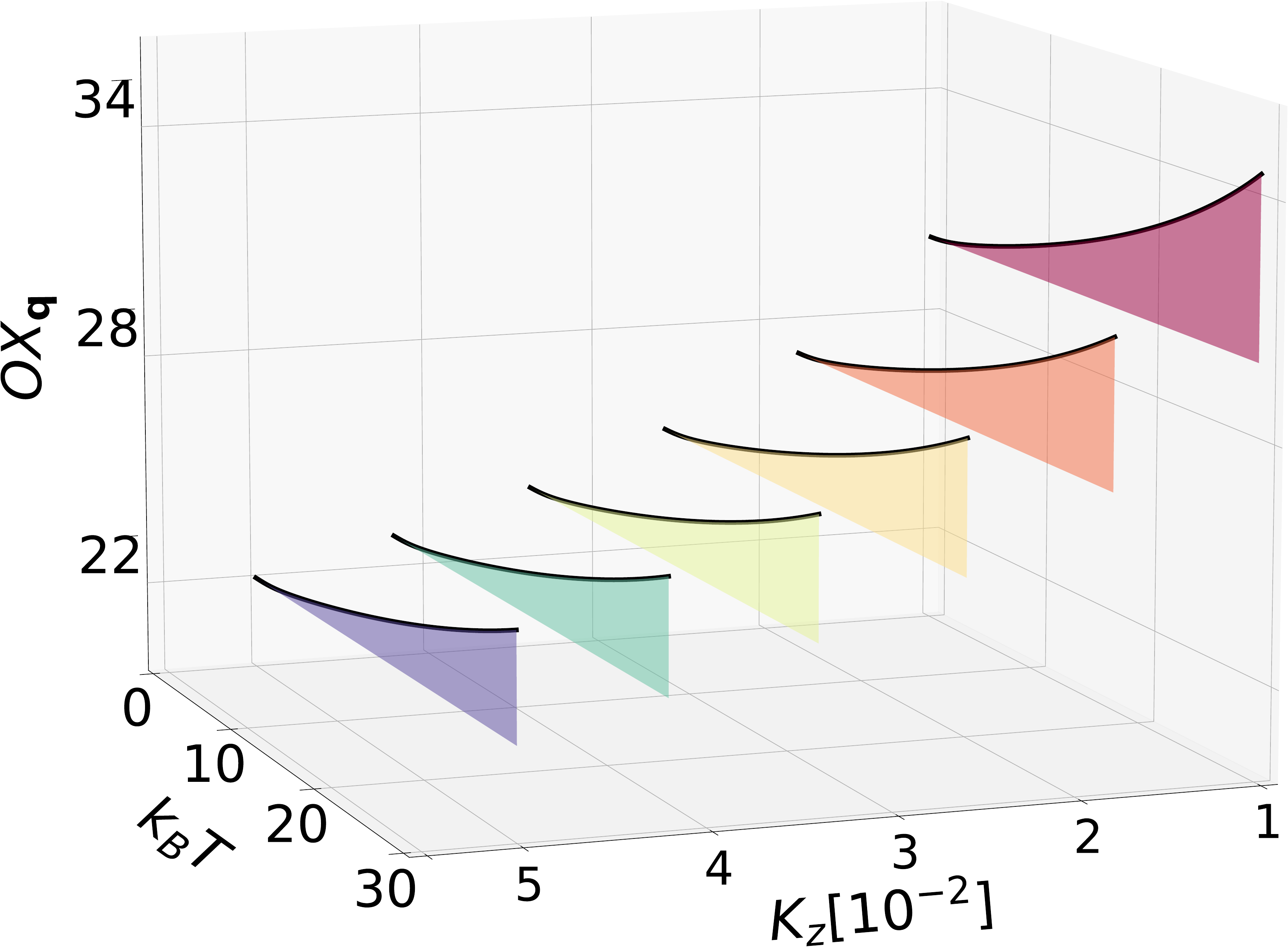}}
    \caption{The squeeze factor $OX_{\mathbf{q}}$ as a function of temperature and anisotropy at $\Gamma$-point.}
    \label{fig:TKSO1}
\end{figure}
Moreover, Fig.\ \ref{fig:SPAQ} demonstrate that temperature enhances the squeezed property of the vacuum ground state $\ket{\psi_{00}^{\mathbf{q}}}$ by further squeezing quantum fluctuations in the position quadrature observable. Conversely, Fig.\ \ref{fig:SPAQ} reveals that anisotropy weakens the squeezed property of the vacuum ground state by stretching quantum fluctuations in the position quadrature observable. This becomes more clear in evaluation of the squeeze factor
\begin{eqnarray}
OX_{\bf q}&=&-10\log\left[\frac{\Delta^{2}X_{\bf q}(\mathcal{K}_z, T)}{\Delta^{2}X_{\bf q}(r_{\mathbf{q}} = 0)}\right]\nonumber\\
&=&-10\log\left[2 \Delta^{2}X_{\bf q}(\mathcal{K}_z, T)\right],
\label{eq:Order-X}
\end{eqnarray}  
as a function of temperature and anisotropy in Fig. \ref{fig:TKSO1}. The squeeze factor $OX_{\mathbf{q}}$ measures a logarithmic growth of the variance of the position quadrature with respect to the vacuum state in $(\eta, \zeta)$ modes, $\Delta^{2}X_{\bf q}(\mathcal{K}_z, T)$, relative to $\Delta^{2}X_{\bf q}(r_{\mathbf{q}} = 0)$, which is the variance of the position quadrature with respect to the vacuum state in the Kittel modes $(a, b)$. The Fig. \ref{fig:TKSO1} shows that the squeeze factor is an increasing function with respect to temperature and a decreasing function with respect to anisotropy.

In Fig.\ \ref{fig:SPAQ}, we also notice a non-trivial conjugate two-mode magnon squeezing effect concerning the quadrature phases $\nu=\pi/2, \pi$. Explicitly, we find that while increasing temperature squeezes (stretches) fluctuations in the position (momentum) observable, the anisotropy does the opposite, namely, increasing anisotropy induces squeezing (stretching) of fluctuations in the momentum (position) observable. In other words, while temperature causes amplitude squeezing, anisotropy causes phase squeezing. This temperature-anisotropy conjugate squeezing is observed in a broader range of parameter space in Fig.\ \ref{fig:SPAQ1}. As shown in Fig. \ref{fig:SPAQ1}, the variance of position quadrature, $\Delta^{2}X_{\mathbf{q}}(\mathcal{K}_z, T)$, decreases with $T$ and increases with $\mathcal{K}_{z}$, while the variance of momentum quadrature, $\Delta^{2}P_{\mathbf{q}}(\mathcal{K}_z, T)$, increases with $T$ and decreases with $\mathcal{K}_z$.
    \label{fig:SPAQ1}
\begin{figure}[h]
    \centering
    {\includegraphics[width=0.49\columnwidth]{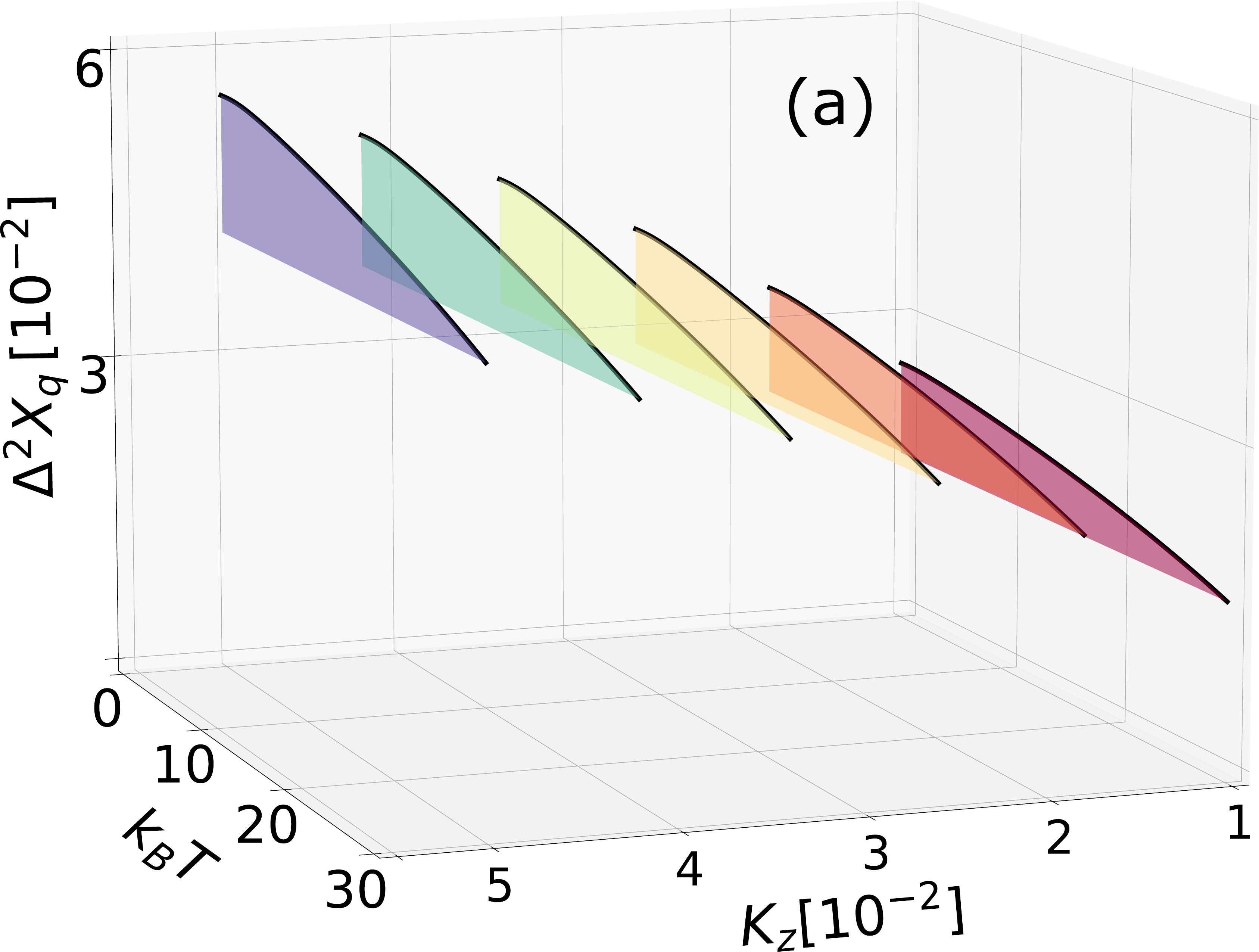}}
    {\includegraphics[width=0.48\columnwidth]{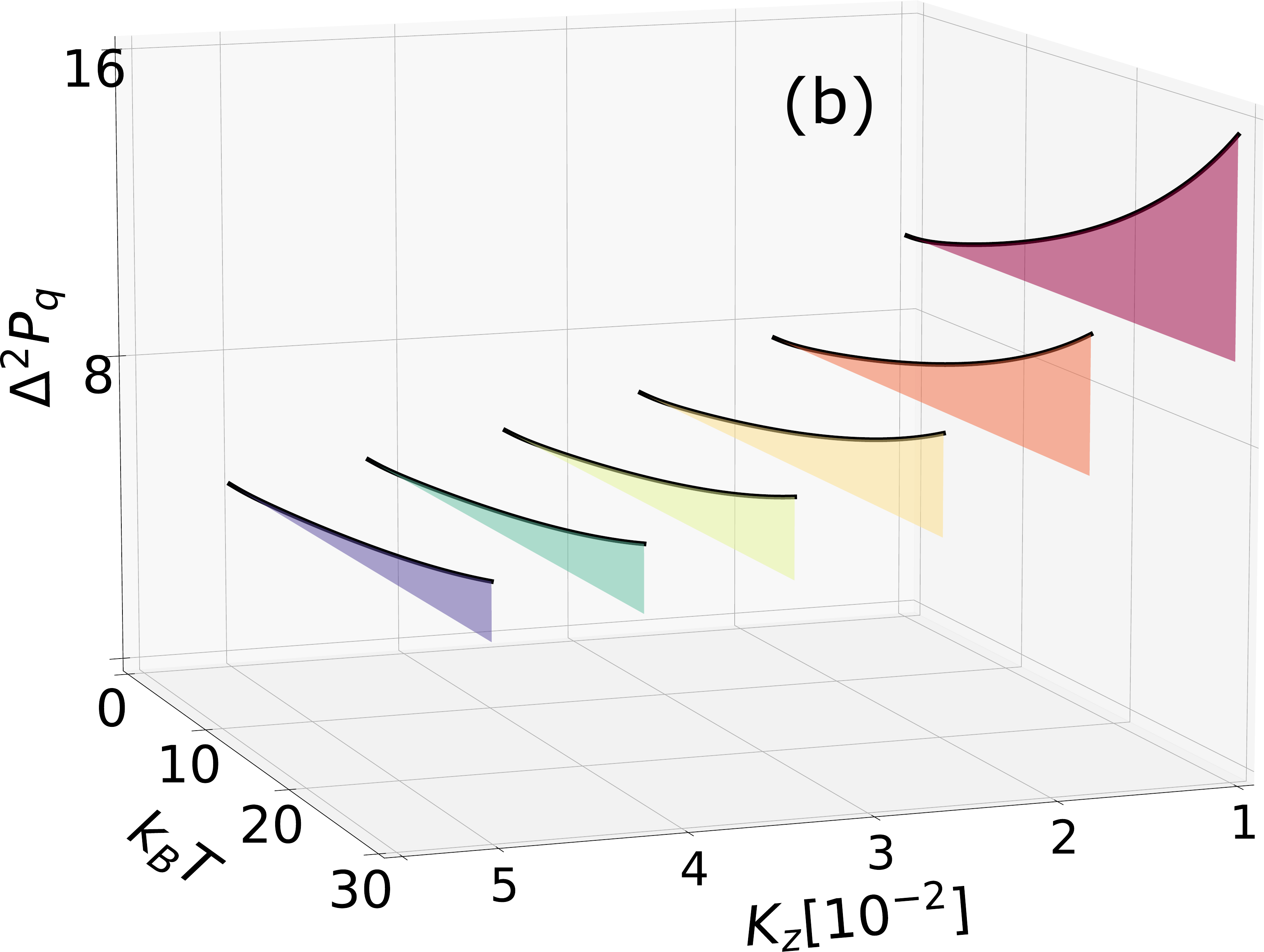}}
    \caption{Quantum fluctuations in the conjugate observables (a) $\Delta^{2}X_{\bf q}$ and (b) $\Delta^{2}P_{\bf q}$ in terms of temperature and uniaxial anisotropy at $\Gamma$-point. While $\Delta^{2}X_{\bf q}$ is a decreasing function of $T$ and an increasing function of  $\mathcal{K}_{z}$, $\Delta^{2}P_{\bf q}$ is an increasing function of $T$ and a decreasing function of $\mathcal{K}_{z}$.}
    \label{fig:SPAQ1}
\end{figure}

\begin{figure}[h]
    \centering
    {\includegraphics[width=0.48\columnwidth]{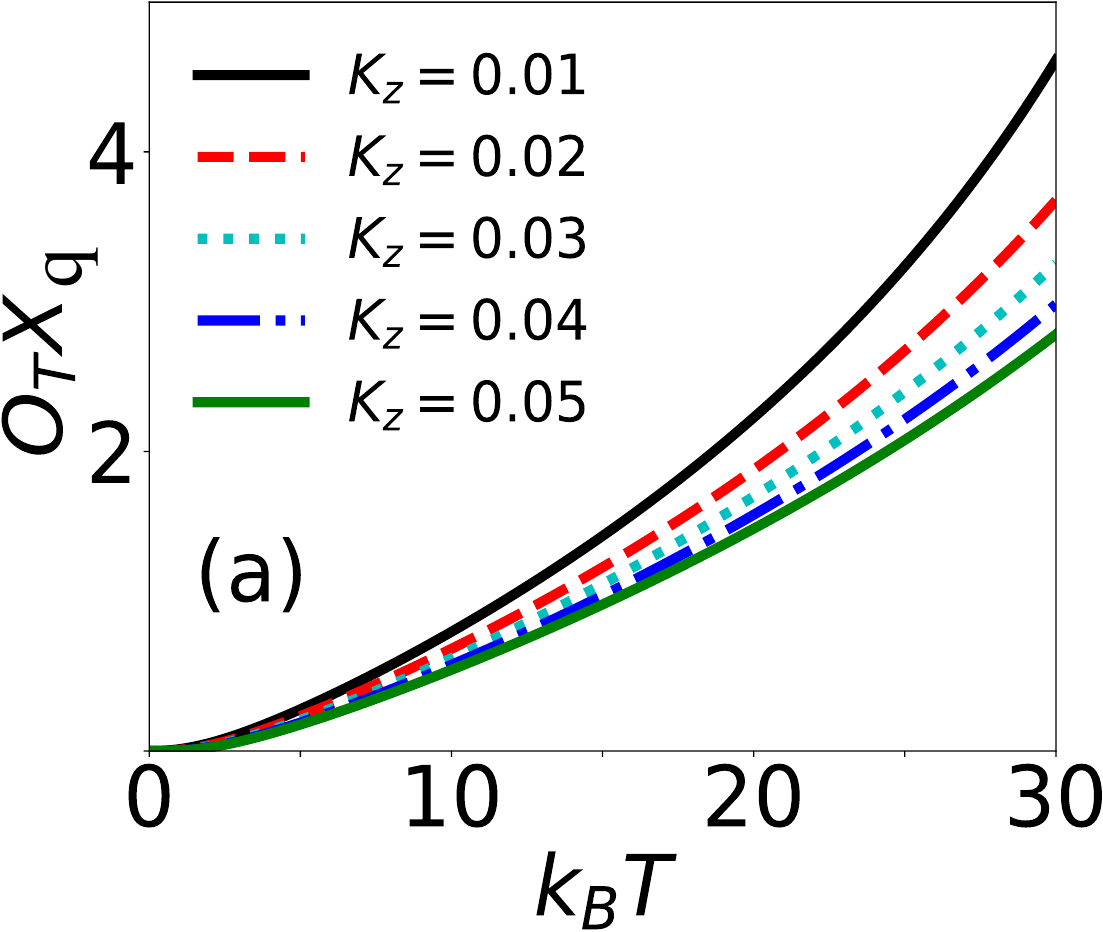}}
    {\includegraphics[width=0.49\columnwidth]{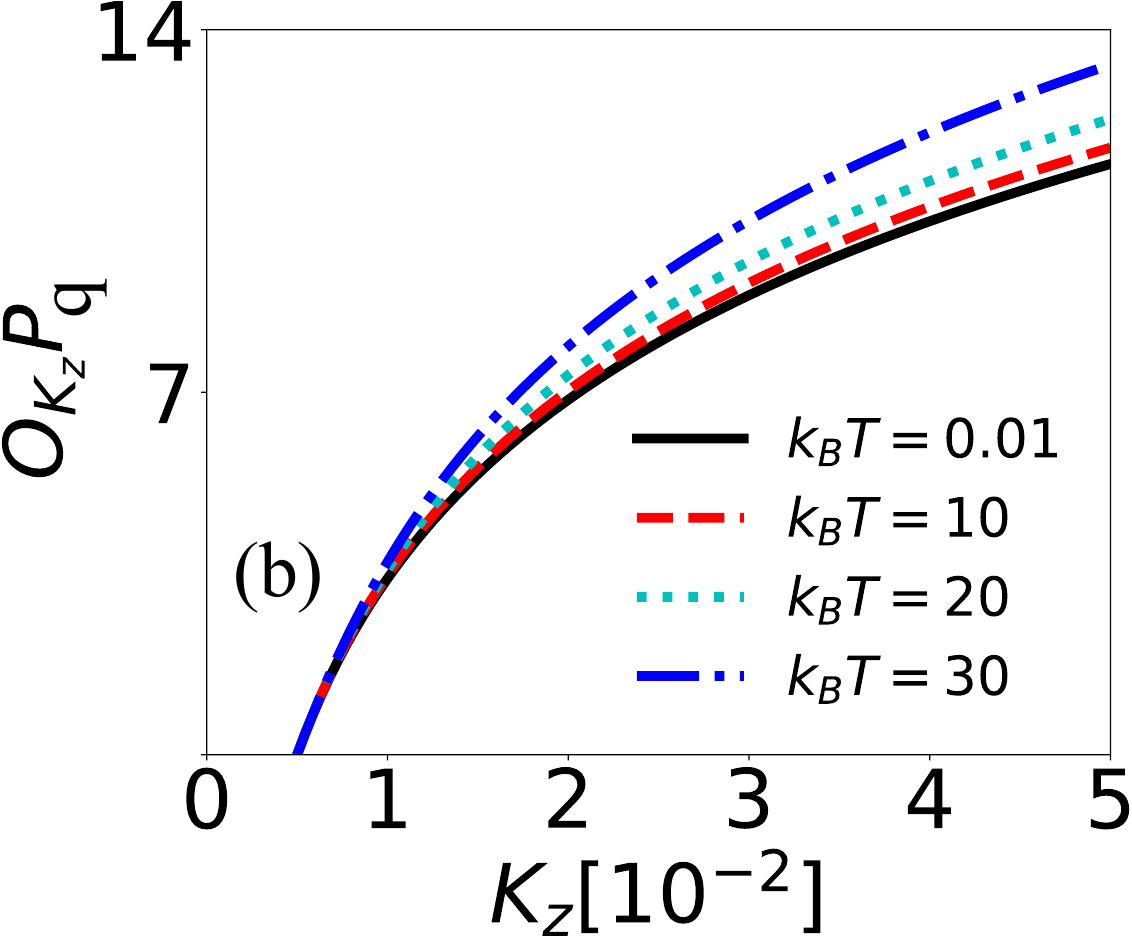}}
    \caption{The squeeze factors (a) $O_{T}X_{\bf q}$ as a function of temperatures for different values of anisotropies, (b) $O_{\mathcal{K}_z}P_{\bf q}$ as a function of anisotropy for different temperatures at $\Gamma$-point.}
    \label{fig:TKSO}
\end{figure}

To further analyze the temperature-anisotropy conjugate two-mode magnon squeezing effect, we introduce the following squeeze factors to quantify the squeezing power induced exclusively by temperature and anisotropy
\begin{eqnarray}
O_{T}X_{\bf q}&=&-10\log\left[\frac{\Delta^{2}X_{\bf q}(\mathcal{K}_z, T)}{\lim_{T \to 0}\Delta^{2}X_{\bf q}(\mathcal{K}_z, T)}\right],\nonumber\\
O_{\mathcal{K}_z}P_{\bf q}&=&-10\log\left[\frac{\Delta^{2}P_{\bf q}(\mathcal{K}_z, T)}{\lim_{\mathcal{K}_z \to 0}\Delta^{2}P_{\bf q}(\mathcal{K}_z, T)}\right]\nonumber\\
\label{eq:Order-para}
\end{eqnarray}
for given $\mathcal{K}_z$ and $T$. For a given value of anisotropy $\mathcal{K}_z$, $O_{T}X_{\bf q}$ quantifies logarithmic growth of the variance of the position quadrature at a finite temperature relative to zero temperature. Similarly, at a given temperature $T$, $O_{\mathcal{K}_z}P_{\bf q}$ quantifies logarithmic growth of the variance of the momentum quadrature for a finite value of anisotropy relative to isotropic case $\mathcal{K}_z \to 0$. Comparing to the total squeeze factor in Eq. \eqref{eq:Order-X}, which takes into account contributions from all Hamiltonian parameters including the antiferromagnetic Heisenberg exchange coupling $J$, the individual squeeze factors in Eq. \eqref{eq:Order-para} evaluates the contribution of only one parameter, either $T$ or $\mathcal{K}_z$, to the squeezing property of the vacuum ground state $\ket{\psi_{00}^{\mathbf{q}}}$. Fig.\ \ref{fig:TKSO} illustrates the squeeze factors in Eq. \eqref{eq:Order-para} corresponding to two conjugate observables. The plots in  Fig.\ \ref{fig:TKSO}, clearly confirms the temperature-anisotropy conjugate two-mode magnon squeezing effect. Moreover, it is evident from the plots that the contribution of anisotropy to the squeezing properties of the system is relatively larger compared to the contribution of temperature. However, it is important to consider that the squeezed nature of the vacuum ground state is associated with the squeezing of quantum fluctuations in the position quadrature. Thus, temperature makes a constructive contribution to the magnon squeezing, while anisotropy has a destructive contribution. This behavior arises due to the conjugate relationship between the position and momentum quadratures.

\begin{figure}
    \centering
    {\includegraphics[width=0.49\columnwidth]{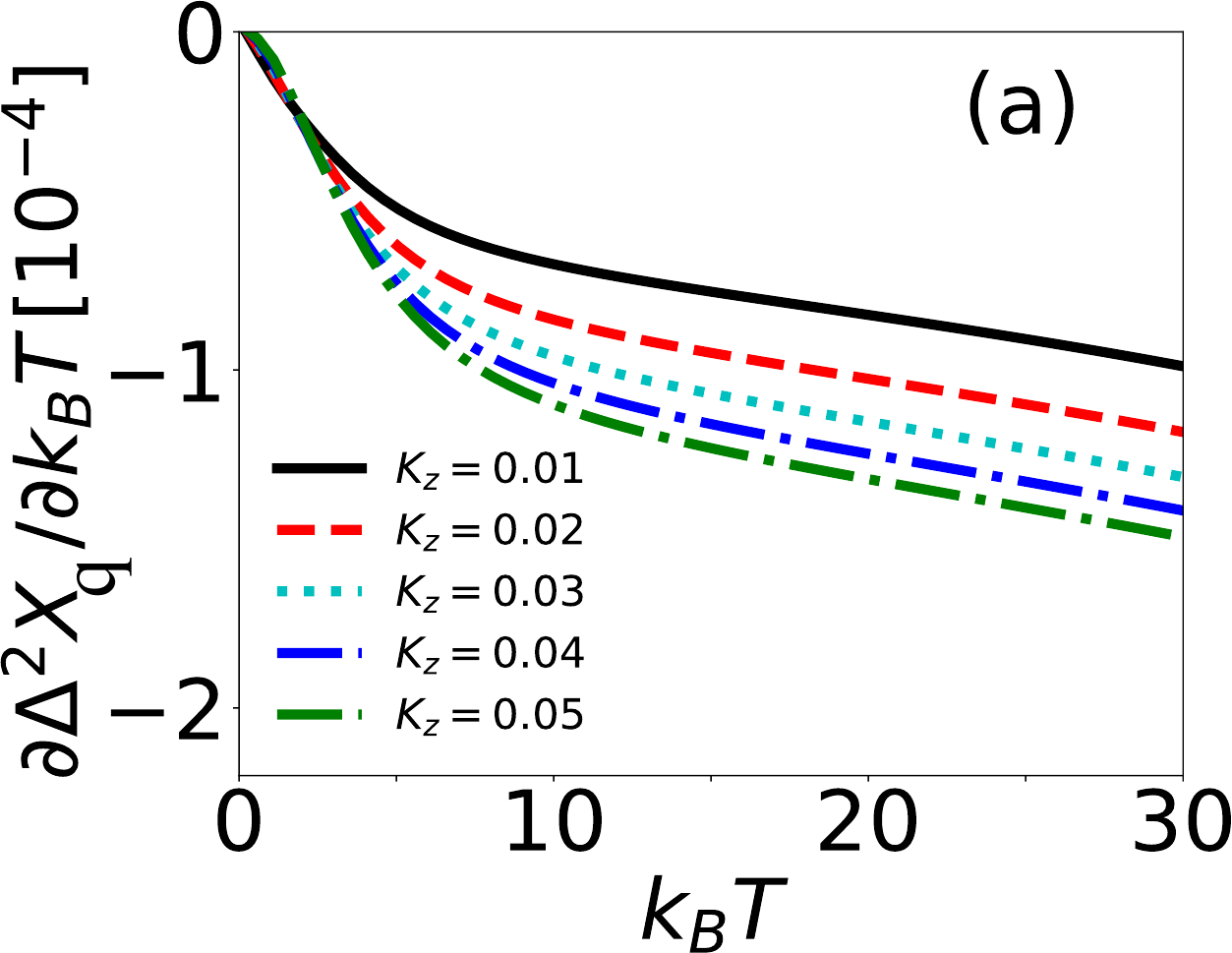}}
    {\includegraphics[width=0.48\columnwidth]{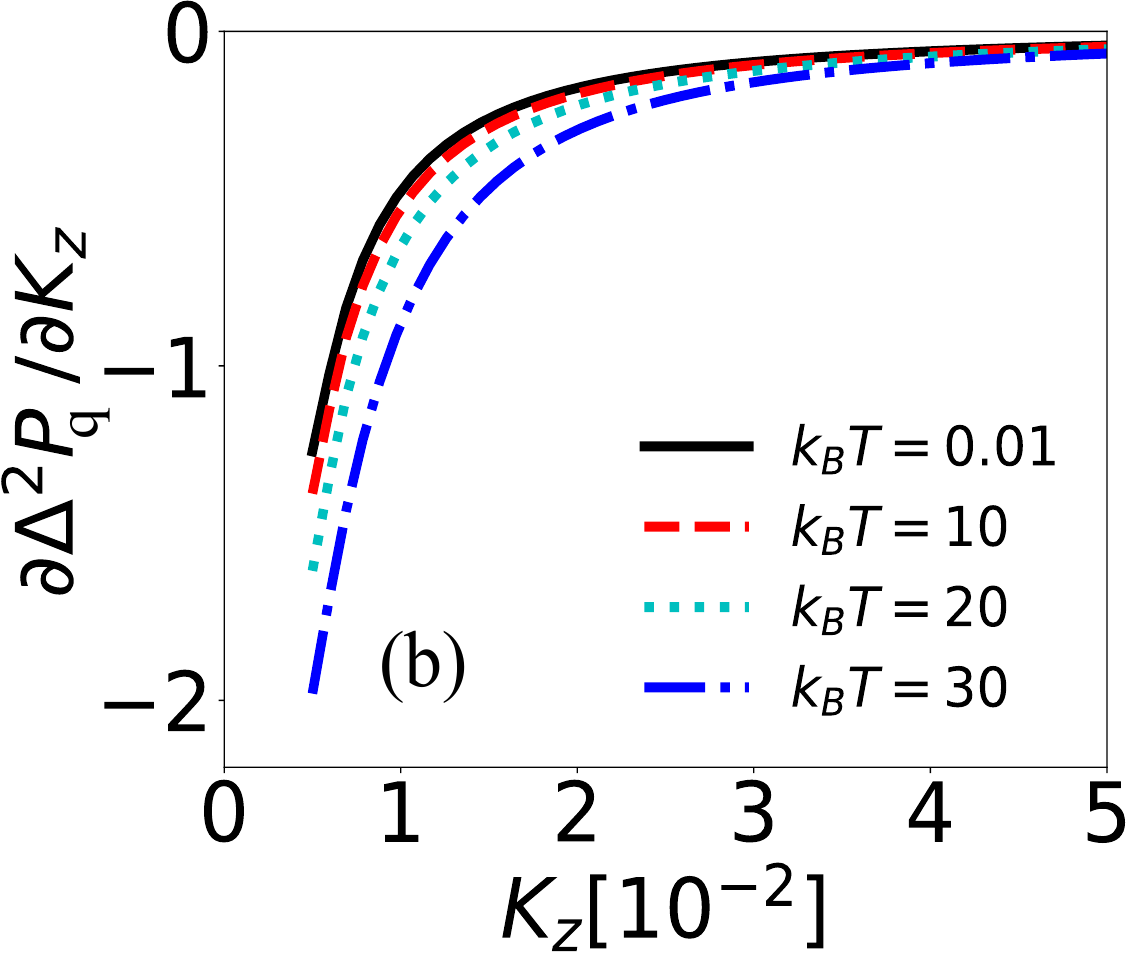}}
    \caption{The rate of change of (a) $\Delta^{2}X_{\bf q}$  in terms of temperature and (b) $\Delta^{2}P_{\bf q}$ in terms of uniaxial anisotropy at $\Gamma$-point. }
    \label{fig:KT-MSR}
\end{figure}

We analyze the rates of anisotropy- and temperature-induced two-mode magnon squeezing by examining the falling slope of quantum fluctuations with respect to temperature and anisotropy in Fig.\ \ref{fig:KT-MSR}. Negative slopes indicate decreasing quantum fluctuations and, consequently, quantum squeezing. It is observed that the rate of squeezing (the negativity of the slope) always increases with temperature. However, the rate for anisotropy-induced squeezing decreases and tends to vanish within a narrow range of uniaxial anisotropy. This indicates that, for a given temperature, the variance $\Delta^{2}P_{\bf q}$ and thus the anisotropy-induced squeeze factor $O_{\mathcal{K}z}P_{\bf q}$ quickly approach constant values after a finite value of anisotropy. In other words, after an initial sharp squeezing effect induced by anisotropy, increasing anisotropy does not contribute further squeezing at any temperature. This is indeed promising, taking into account the competition between temperature and anisotropy in two-mode magnon squeezing and the destructive effect of anisotropy on the squeezing properties of the system.

\begin{figure}[h]
    \centering  
    {\includegraphics[width=0.48\columnwidth]{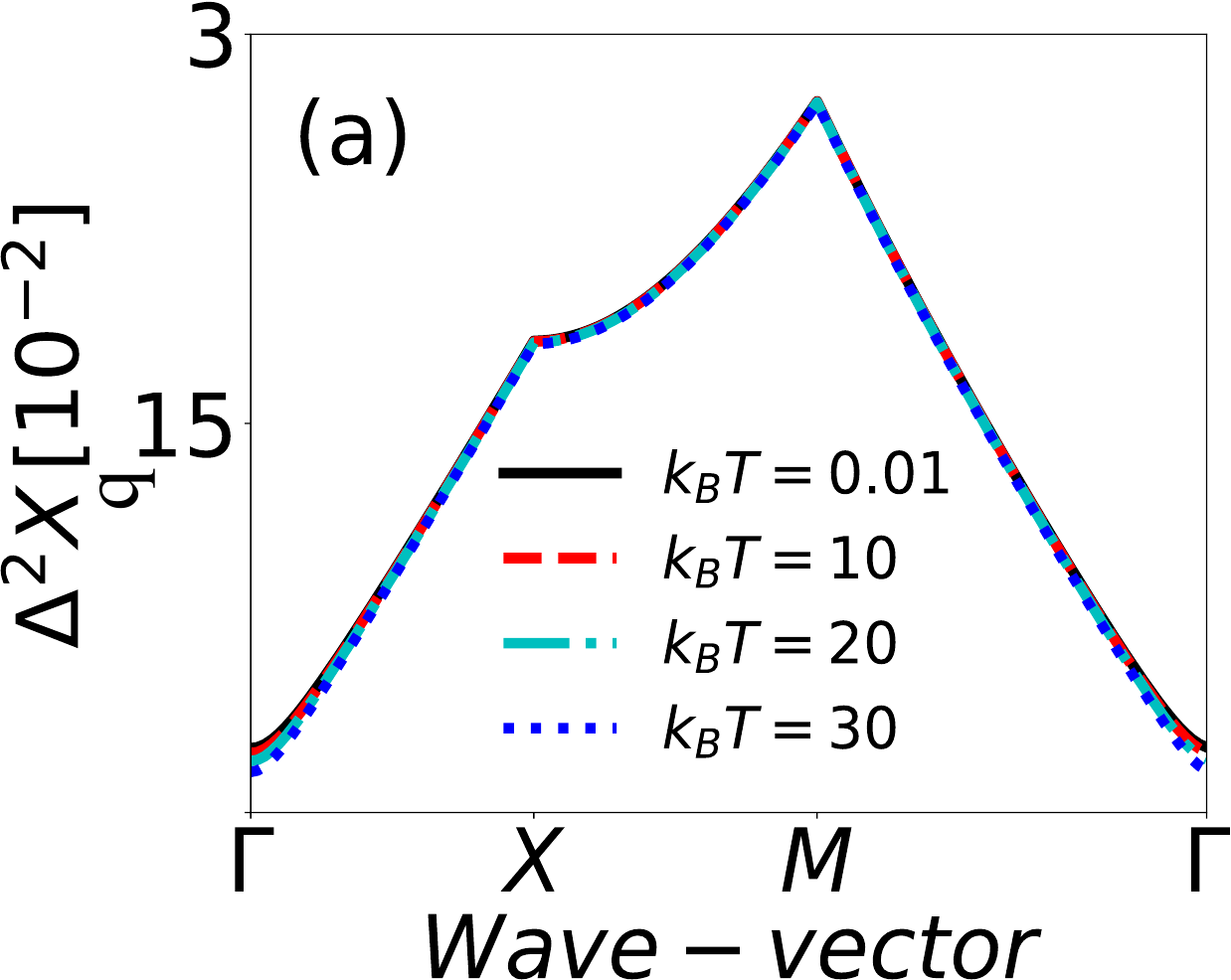}}
    {\includegraphics[width=0.48\columnwidth]{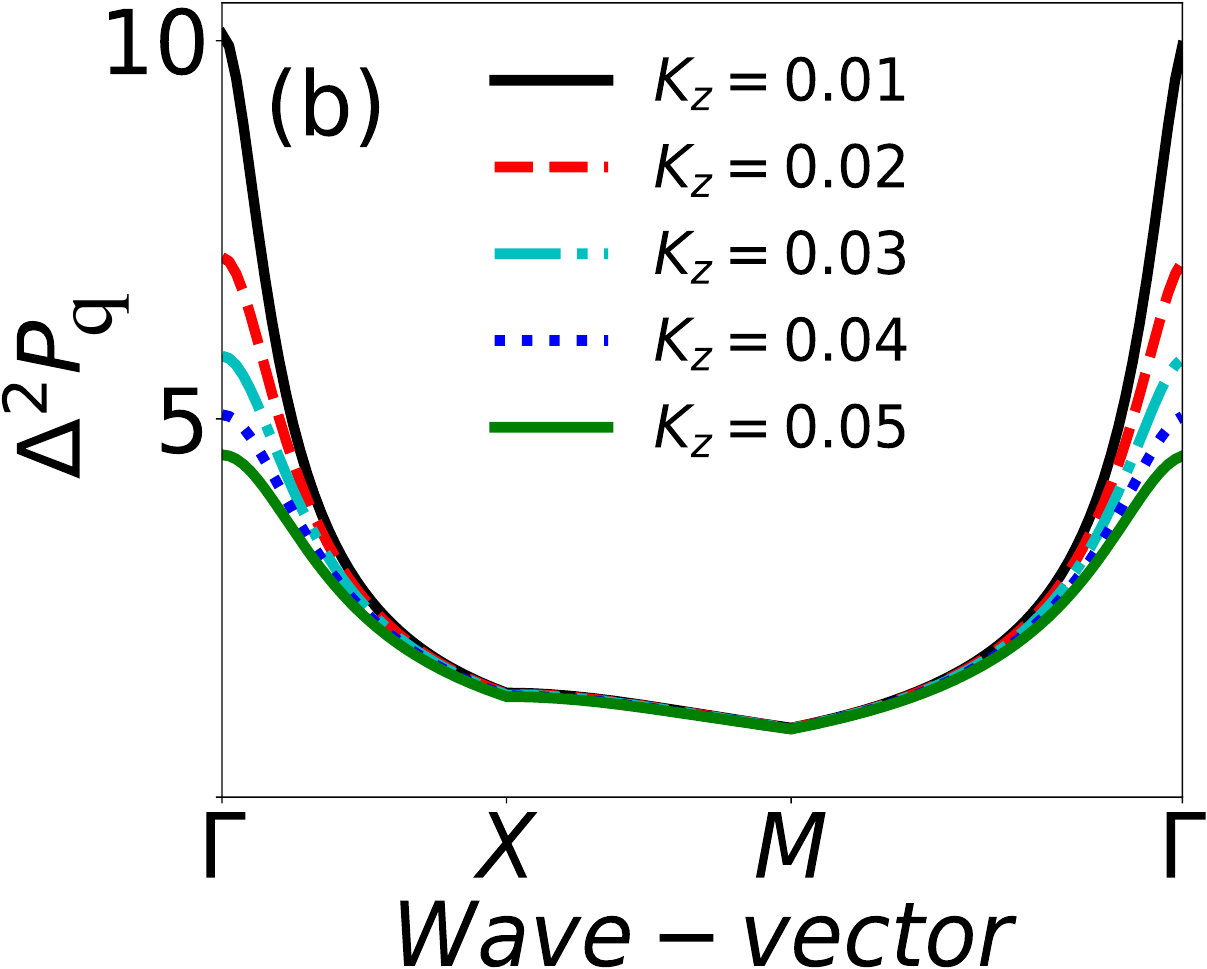}}
    {\includegraphics[width=0.48\columnwidth]{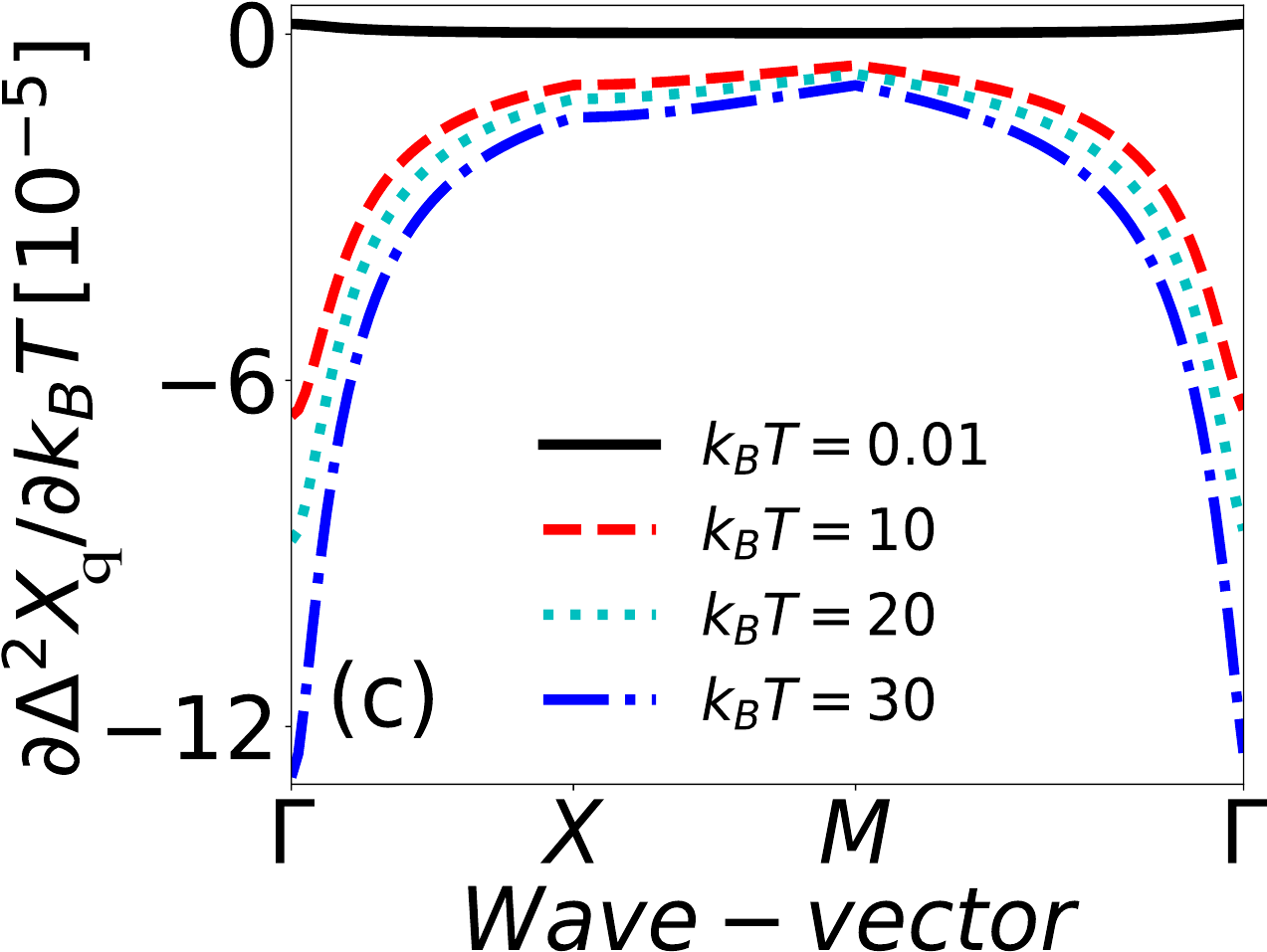}}
    {\includegraphics[width=0.48\columnwidth]{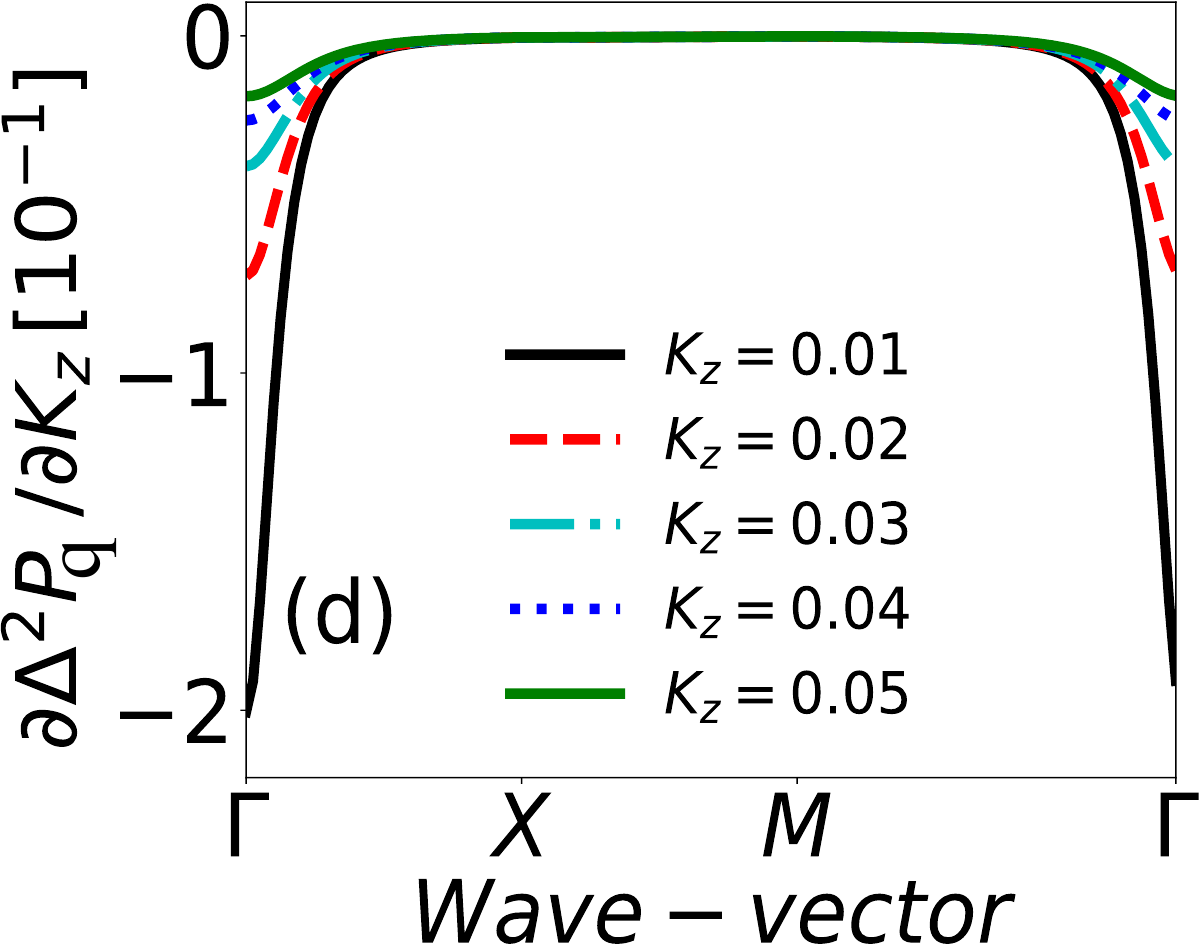}}
    \caption{Quantum fluctuations and their rate of changes at different $\mathbf{q}$ points along the high symmetric path in the Brillouin zone of a square lattice. In panels (a, c) $K_z = 0.01$ meV and panels (b, d) $k_BT = 1$ meV. The most pronounced squeezing effects occur at the zone center.}
    \label{fig:dispersion3}
\end{figure}

 We end our discussion with a few remarks. Firstly, temperature-anisotropy conjugate magnon squeezing, which was observed above at the Brillouin zone center, occurs for any point in the Brillouin zone. To confirm this point, the corresponding squeezings along the high-symmetric path in the Brillouin zone associated with a square lattice are illustrated in Fig.\ \ref{fig:dispersion3}. As show in this figure, for each point in the Brillouin zone, temperature induces squeezing of quantum fluctuations in the position quadrature (amplitude squeezing) while anisotropy induces squeezing of quantum fluctuations in the conjugate momentum quadrature (phase squeezing). Despite the fact that magnon squeezing occurs everywhere in the Brillouin zone, the strongest temperature-anisotropy squeezing effects are at the center of the Brillouin zone. This squeezing property, combined with other characteristics of the zone center magnons, such as low-energy magnons (Fig. \ref{fig:dispersion1}), stability at higher anisotropy (Fig. \ref{fig:dispersion2}), and strong two-mode quantum correlation \cite{PhysRevB.102.224418, PhysRevB.104.224302}, renders zone center magnons in antiferromagnetic materials particularly intriguing for quantum magnonics and its potential applications in sustainable quantum technologies.

\begin{figure}[h]
    \centering
    {\includegraphics[width=0.48\columnwidth]{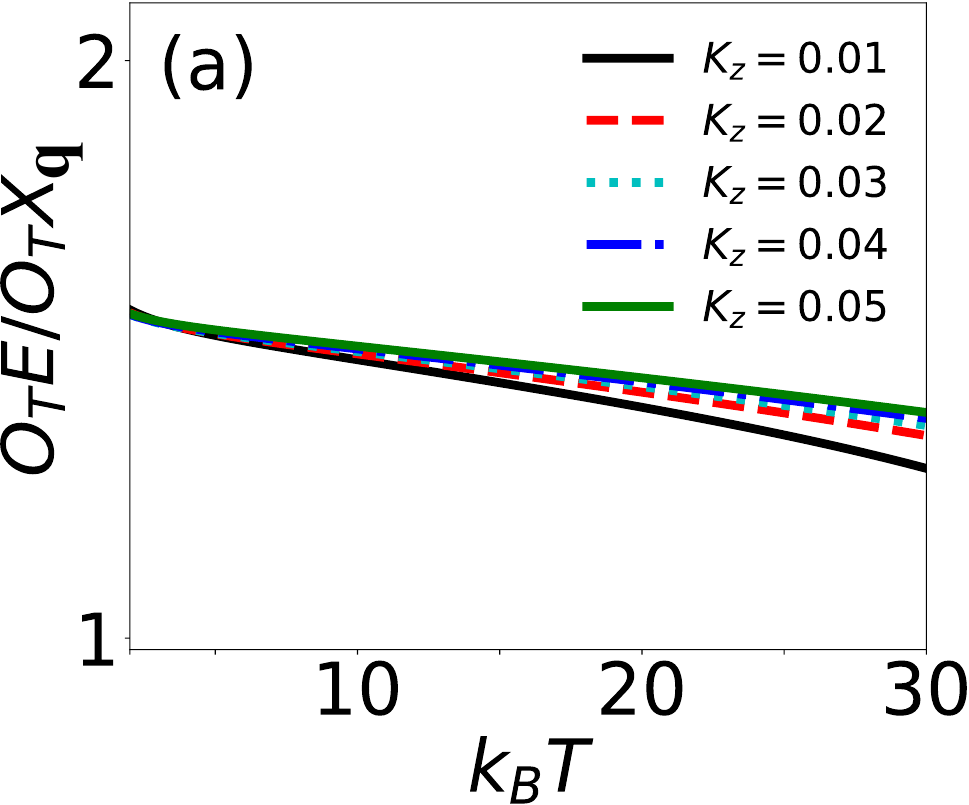}}
    {\includegraphics[width=0.49\columnwidth]{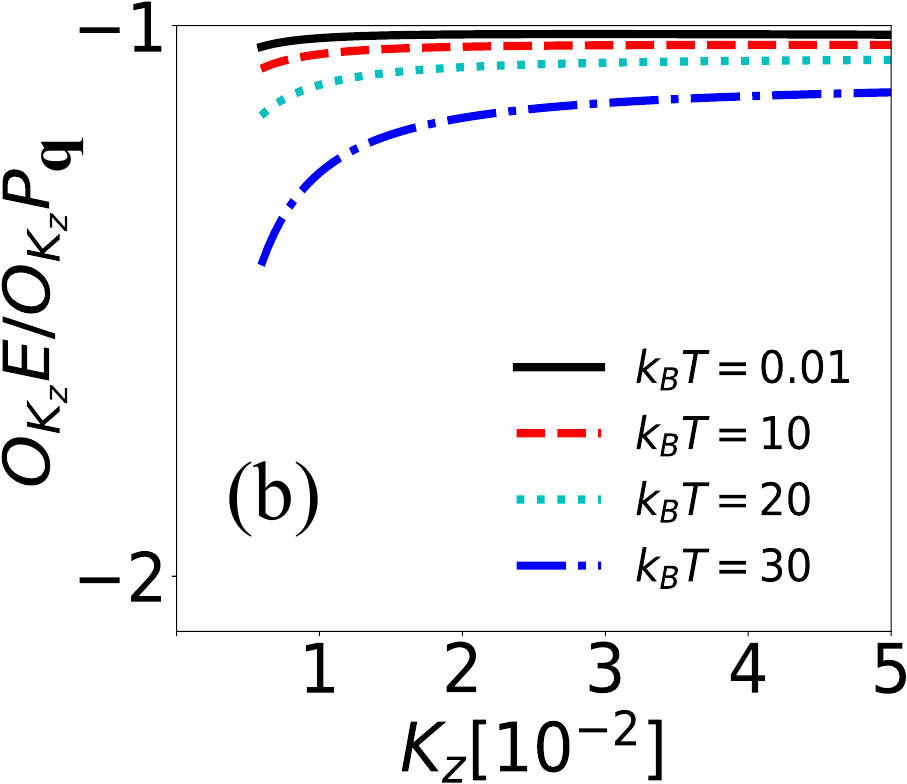}}
    \caption{The correlation between squeezing and the variation of energy with respect to (a) temperature and (b) uniaxial anisotropy, respectively, for discrete values of anisotropy and temperature at $\Gamma$-point. In both cases, there is a linear correlation between the factors of squeezing and change of energy except for very small anisotropies. The positive correlation in the panel (a) indicates that the energy $\mathcal{E}_{\bf q}$ and the quantum fluctuation $\Delta^{2}X_{\bf q}$ are both decreasing functions of $T$. The negative correlation in the panel (b) shows that although increasing $\mathcal{K}_z$ reduces the quantum fluctuation $\Delta^{2}P_{\bf q}$ the energy $\mathcal{E}_{\bf q}$ is an increasing function of $\mathcal{K}_z$.} 
    \label{fig:TKSOEO}
\end{figure}

Secondly, as shown in Figs. \ref{fig:dispersion1} and \ref{fig:dispersion2}, temperature and anisotropy affect the energy of the system in a similar manner that they effect two-mode magnon squeezing. While temperature decreases the energy of the system, anisotropy increases the energy. In order to compare the effects of temperature and anisotropy on energy and squeezing, we introduce the factors of changes in energy induced by temperature and anisotropy 
\begin{eqnarray}
O_{T}E_{\bf q}&=&-10\log\left[\frac{\mathcal{E}_{\bf q}(\mathcal{K}_z, T)}{\lim_{T \to 0}\mathcal{E}_{\bf q}(\mathcal{K}_z, T)}\right],\nonumber\\
O_{\mathcal{K}_z}E_{\bf q}&=&-10\log\left[\frac{\mathcal{E}_{\bf q}(\mathcal{K}_z, T)}{\lim_{\mathcal{K}_z \to 0}\mathcal{E}_{\bf q}(\mathcal{K}_z, T)}\right],\nonumber\\
\label{eq:Order-para-E}
\end{eqnarray}
equivalent to the squeeze factors defined in Eq. \eqref{eq:Order-para}.
For a given value of anisotropy $\mathcal{K}_z$, $O_{T}E_{\bf q}$ quantifies logarithmic growth of the desperation energy at a finite temperature relative to zero temperature. Similarly, at a given temperature $T$, $O_{\mathcal{K}_z}E_{\bf q}$ quantifies logarithmic growth of the desperation energy for a finite value of anisotropy relative to isotropic case $\mathcal{K}_z \to 0$.
We examine the correlation between squeeze factors and the factor of changes in energy caused by temperature and anisotropy in Fig.\ \ref{fig:TKSOEO}. In both cases, we observe a linear correlation, except for very small anisotropy, where antiferromagnetic structures are, in general, not stable. 

Thirdly, our analysis in the present paper, particularly the values of squeeze factors shown in Figs. \ref{fig:TKSO1} and \ref{fig:TKSO}, and their relation to desperation energies mentioned in the previous remarks, indicates that magnons in antiferromagnetic materials provide a promising resource for low-energy stabilized continuous variable bosonic modes, which exhibit an intrinsic and significantly strong two-mode squeezing property, compared to photons in quantum optics.

Last but not least, although in the above analysis, we only focus on magnon squeezing in the vacuum ground state, we notice that a similar scenario holds for all the energy eigenbasis state $\ket{\psi^{{\bf q}}_{nm}}$ given in Eq. \ref{EES}. We provide some details about squeezing in excited states in the Appendix. 

\section{conclusion}
\label{conclusion}
In summary, we consider easy-axis antiferromagnetic materials beyond the linear spin wave theory. We examine the effects of temperature and anisotropy on two-mode magnon squeezing in such materials. As a result of the nonlinearity, we observe a conjugate squeezing effect driven by temperature or anisotropy. Specifically, we demonstrate that temperature gives rise to magnon amplitude squeezing, which refers to the squeezing of quantum fluctuations in the magnon position quadrature. On the other hand, anisotropy leads to magnon phase squeezing, which corresponds to the squeezing of quantum fluctuations in the magnon momentum quadrature. Furthermore, we find that the squeezing nature of the two-mode magnon eigenenergy states is associated with amplitude squeezing rather than phase squeezing. These findings indicate a competition between temperature and anisotropy in a sense that temperature has a constructive impact, while anisotropy has a destructive impact on the squeezing nature of the two-mode magnon eigenenergy states. However, the destructive effect of anisotropy on two-mode magnon squeezing is shown to be bounded by demonstrating that the rate of phase squeezing with respect to anisotropy tends to vanish after a finite value of anisotropy. 
In addition, we explore the correlations between temperature- and anisotropy-induced magnon squeezings and the variations of energy caused by temperature and anisotropy. We observe linear correlations, except for cases of very small anisotropy, where antiferromagnetic systems are known to display some instability especially in lower dimensions. 

Although our analysis and results emerge at any point in the Brillouin zone, enhanced magnon squeezing is predominantly observed at the Brillouin zone center.
Our findings highlight the influential factors that determine higher magnon squeeze factors in uniaxial antiferromagnetic materials. Notably, elevated temperatures, reduced anisotropy levels, and proximity to the center of the Brillouin zone are identified as key contributors to the augmentation of magnon squeezing. These characteristics specifically determine low-energy magnons in the mentioned materials.

\section{acknowledgements}
M. Sh. thanks Stiftelsen Olle Engkvist Byggm\"astare. V.A.M. acknowledges financial support from Knut and Alice Wallenberg Foundation through Grant No. 2018.0060.

\section{Appendix}
\label{Appendix}
\begin{figure}[h]
    \centering
    {\includegraphics[width=0.45\columnwidth]{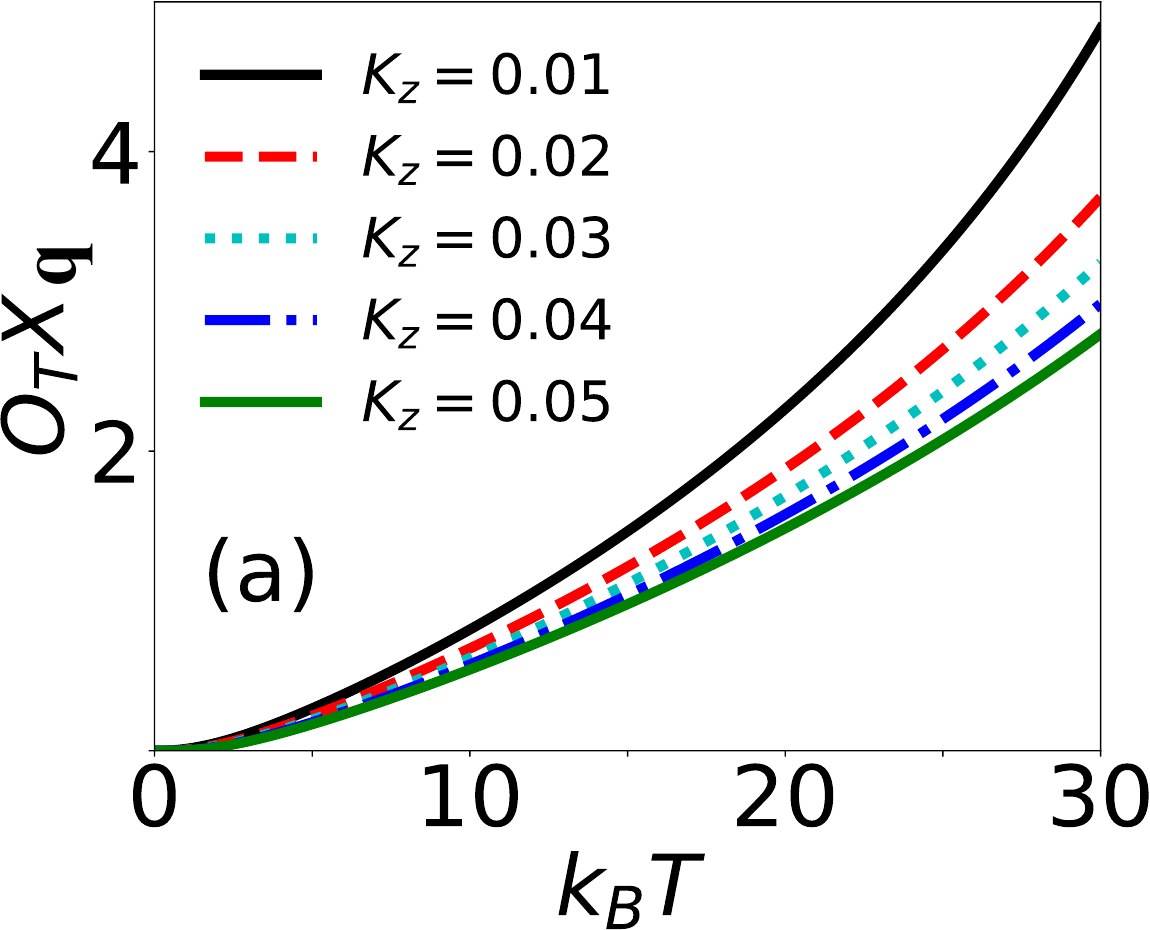}}
    {\includegraphics[width=0.48\columnwidth]{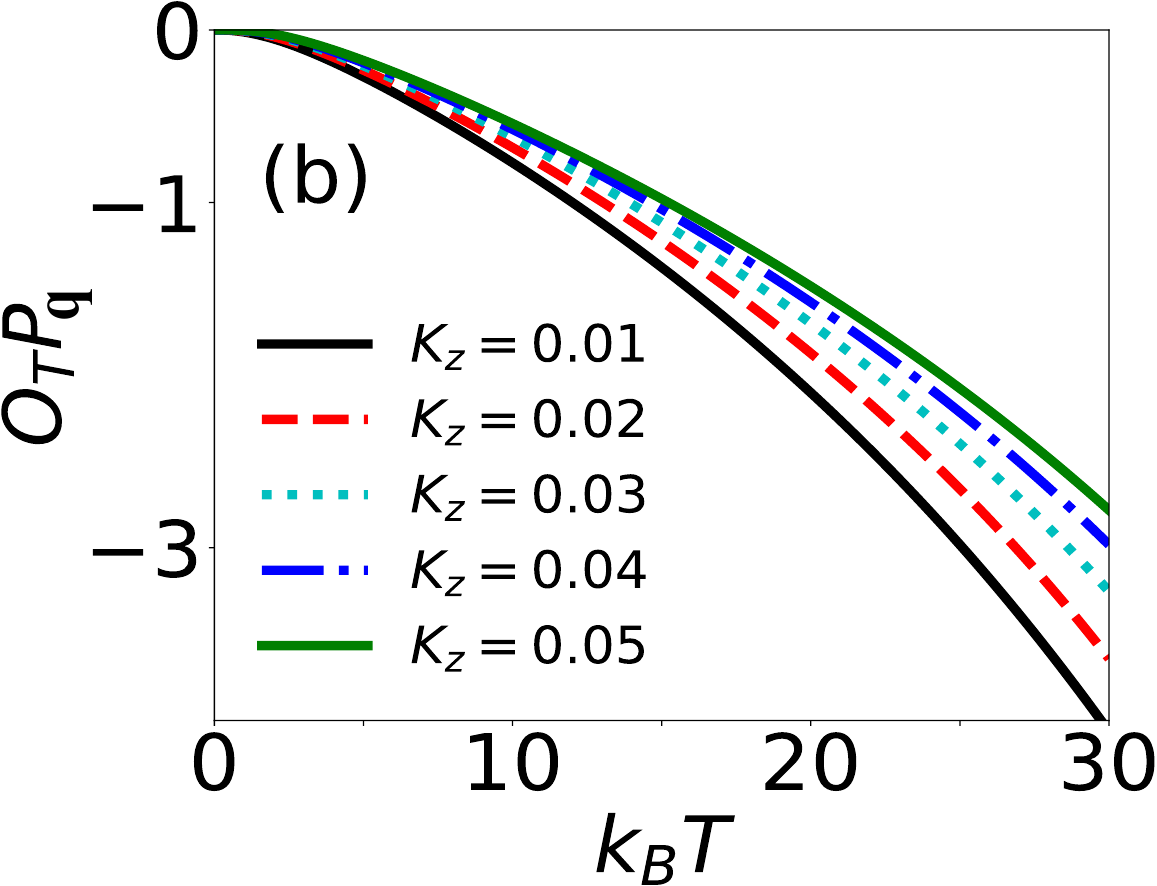}}
    {\includegraphics[width=0.48\columnwidth]{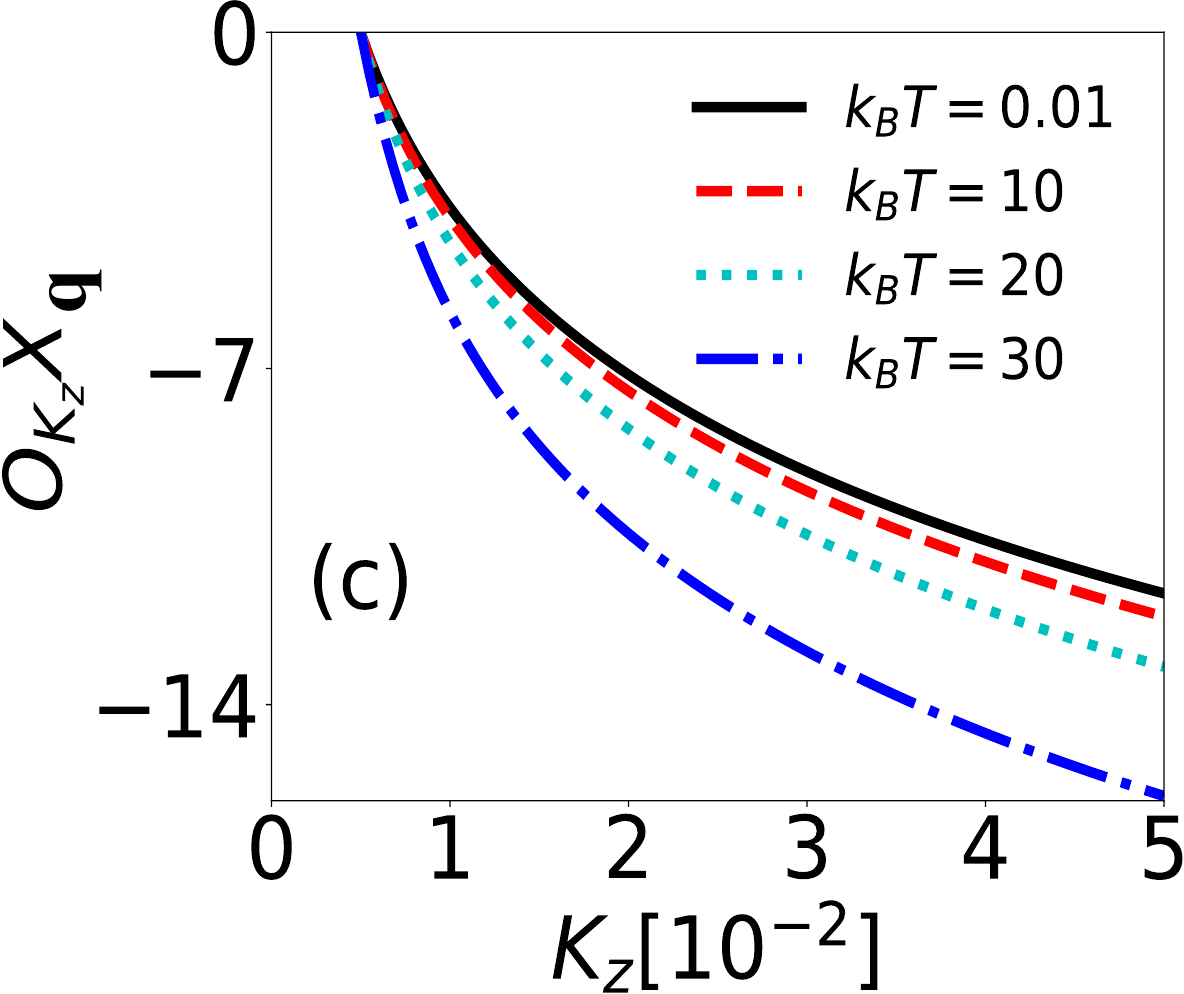}}
    {\includegraphics[width=0.48\columnwidth]{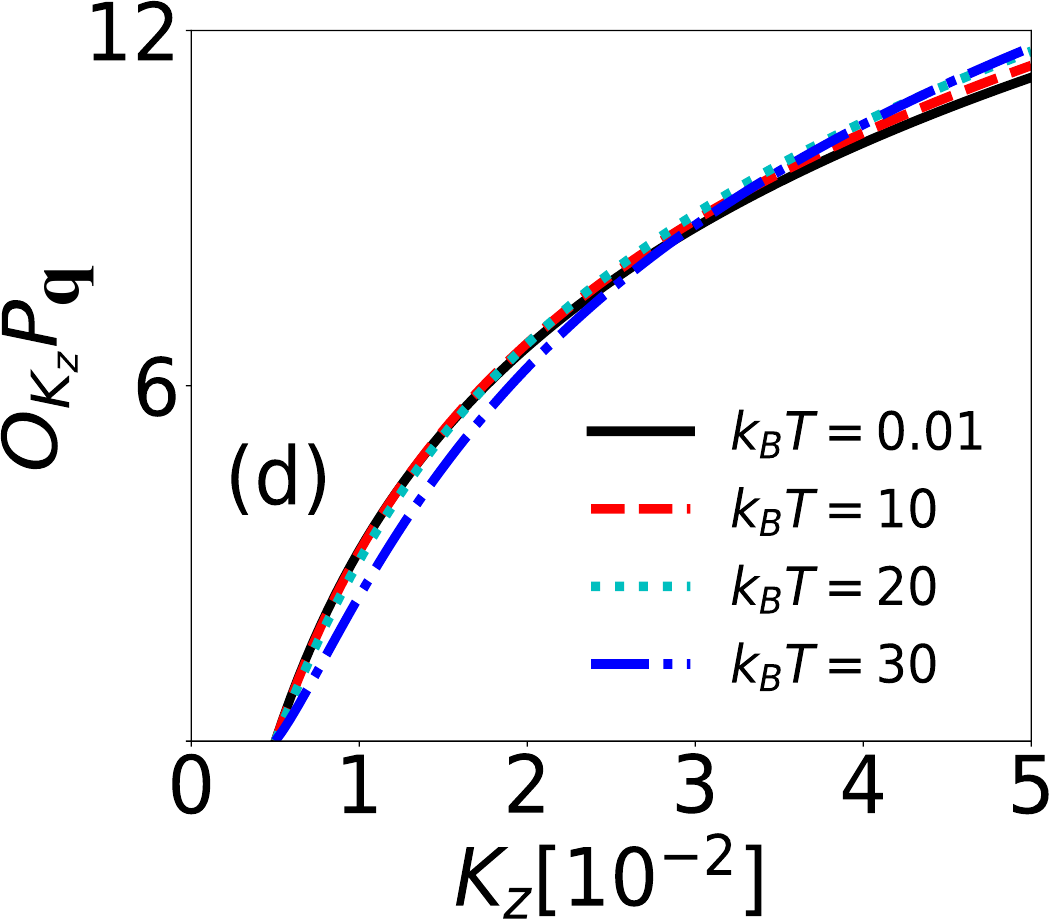}}
    \caption{The squeeze (stretch) factors as a function of anisotropy and temperature for excited states $\ket{\psi^{{\bf q}}_{10}}$ and $\ket{\psi^{{\bf q}}_{01}}$ at $\Gamma$-point. The positive valued factors in panels (a) and (d) indicate squeezing and the negative valued factors in panels (c) and (b) show stretching. Anisotropy and temperature induce squeezing associated with conjugate observables.}
    \label{fig6:n1m0}
\end{figure}

\begin{figure}[h]
    \centering
    {\includegraphics[width=0.45\columnwidth]{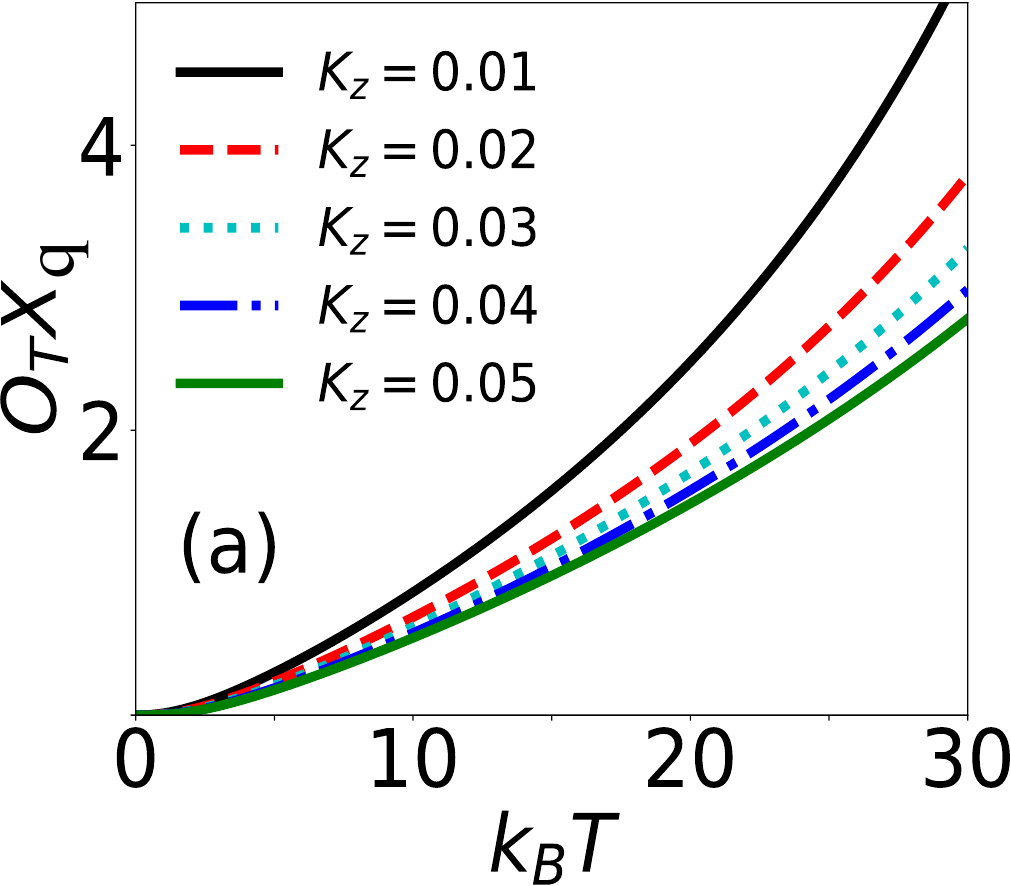}}
    {\includegraphics[width=0.48\columnwidth]{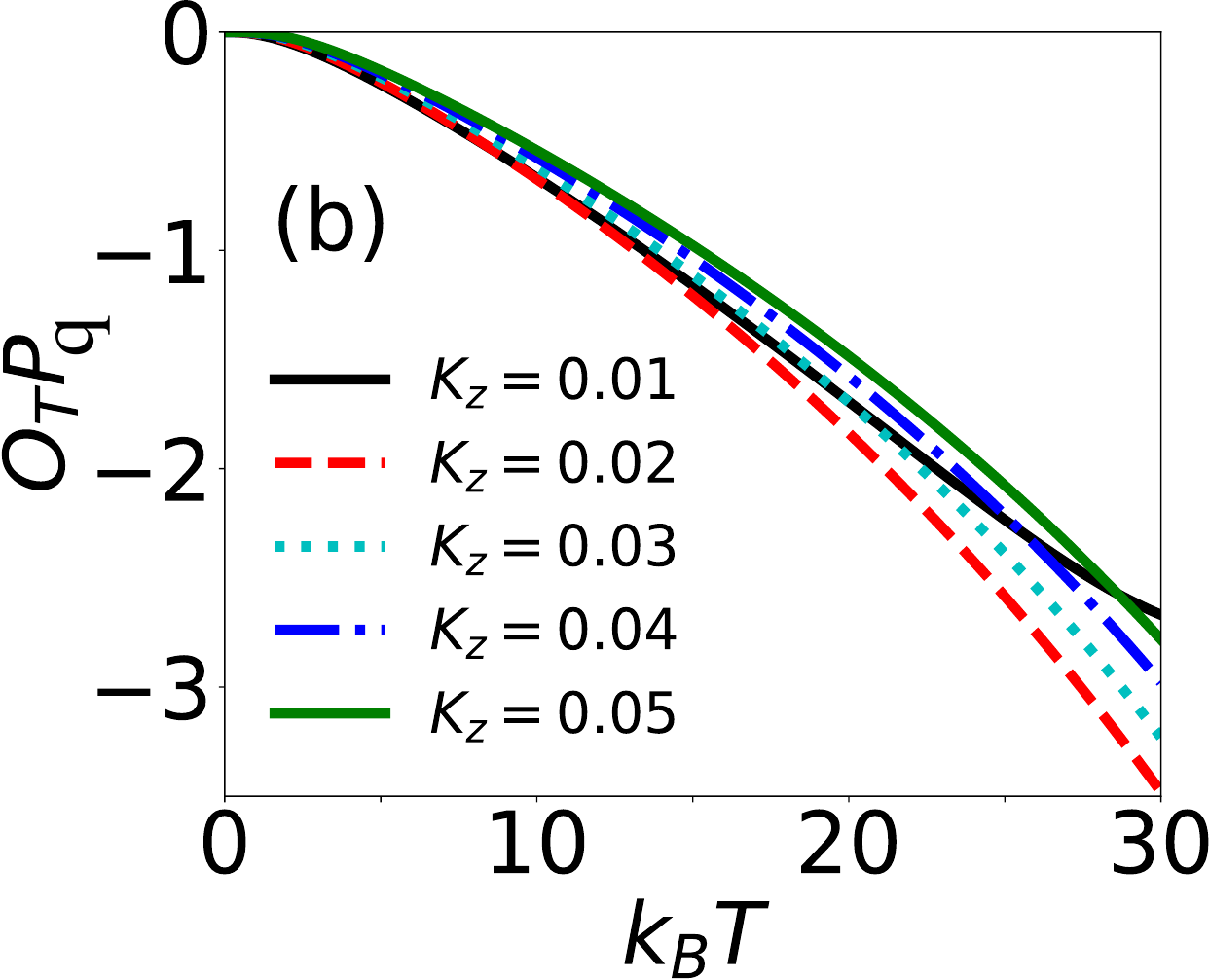}}
    {\includegraphics[width=0.48\columnwidth]{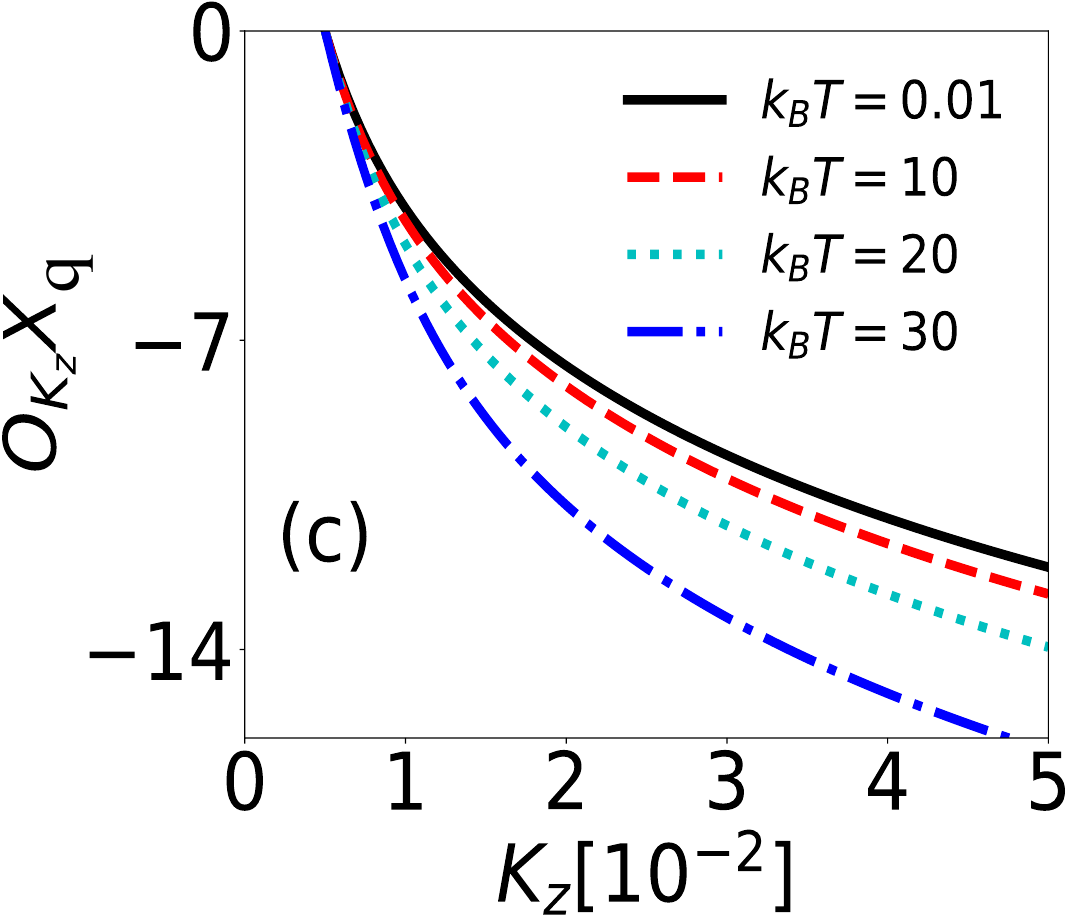}}
    {\includegraphics[width=0.48\columnwidth]{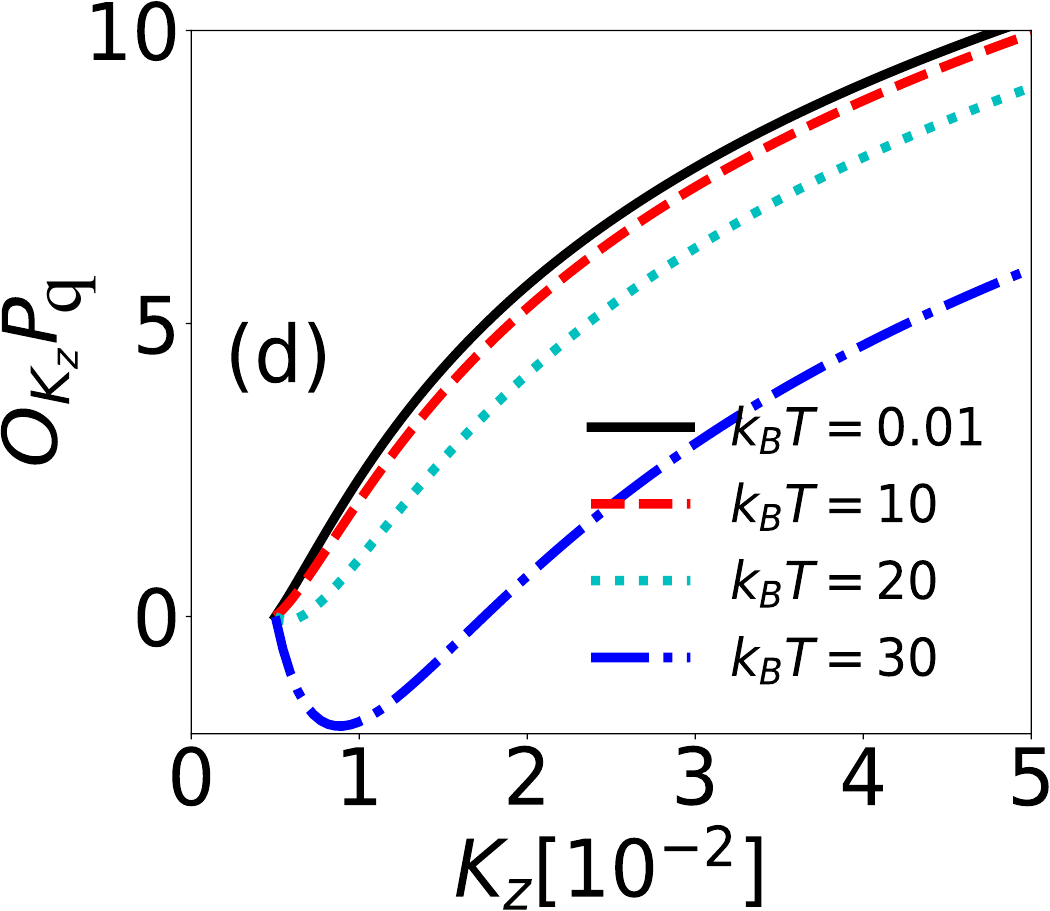}}
    \caption{The squeeze (stretch) factors as a function of anisotropy and temperature for excited states $\ket{\psi^{{\bf q}}_{11}}$ at $\Gamma$-point. The positive valued factors in panels (a) and (d) indicate squeezing and the negative valued factors in panels (c) and (b) show stretching. Anisotropy and temperature induce squeezing associated with conjugate observables.}
    \label{fig6:n1m1}
\end{figure}

The temperature-anisotropy conjugate magnon squeezing effect discussed above is valid for all excited states $\ket{\psi^{{\bf q}}_{nm}}$. To see this, we plot temperature- and anisotropy-induced squeeze factors for a few excited states including $\ket{\psi^{{\bf q}}_{10}}$ and $\ket{\psi^{{\bf q}}_{01}}$ in Fig. \ref{fig6:n1m0}, and $\ket{\psi^{{\bf q}}_{11}}$ in Fig.\ \ref{fig6:n1m1}. In addition to the squeeze factors, we also plot the following complementary factors 
\begin{eqnarray}
O_{\mathcal{K}_z}X_{\bf q}&=&-10\log\left[\frac{\Delta^{2}X_{\bf q}(\mathcal{K}_z, T)}{\lim_{\mathcal{K}_z \to 0}\Delta^{2}X_{\bf q}(\mathcal{K}_z, T)}\right],\nonumber\\
O_{T}P_{\bf q}&=&-10\log\left[\frac{\Delta^{2}P_{\bf q}(\mathcal{K}_z, T)}{\lim_{T \to 0}\Delta^{2}P_{\bf q}(\mathcal{K}_z, T)}\right],\nonumber\\
\label{eq:Order-para-st}
\end{eqnarray}
which are defined similar to the squeeze factors given in Eq. \eqref{eq:Order-para}. While positive values for the factors in Eq. \eqref{eq:Order-para} is indication of squeezing the negative values for the factors in Eq. \eqref{eq:Order-para-st} in the following plots indicates stretch of quantum fluctuation with respect to temperature and anisotropy. These four factors in Eqs. \eqref{eq:Order-para} and \eqref{eq:Order-para-st} clearly captures the temperature-anisotropy conjugate magnon squeezing effect in the following plots. 
\bibliography{ref.bib}
\end{document}